\documentclass[12pt]{article}
\usepackage{amsmath}
\usepackage{longtable}
\usepackage{graphicx}
\usepackage{xcolor}
\usepackage{siunitx}
\usepackage[table]{xcolor}
\usepackage{geometry}
\usepackage{float}
\usepackage{subcaption}
\usepackage[utf8]{inputenc}
\usepackage[T1]{fontenc}
\usepackage{url}
\usepackage{caption}
\usepackage{hyperref}
\usepackage[backend=biber,style=apa]{biblatex}
\title{Integrating Mechanistic Modeling and Machine Learning to Study CD4+/CD8+ CAR-T Cell Dynamics with Tumor Antigen Regulation}
\author{Saranya Varakunan, Melissa Stadt, Mohammad Kohandel \\
Department of Applied Mathematics, University of Waterloo, \\
Waterloo, Ontario, Canada, N2L 3G1}
\date{}

\begin{document}

\begin{titlepage}
  \centering
  \vspace*{0.18\textheight}

  {\LARGE \bfseries Integrating Mechanistic Modeling and Machine Learning to Study CD4$^{+}$/CD8$^{+}$ CAR-T Cell Dynamics
  with Tumor Antigen Regulation\par}

  \vspace{4em}
  {\large
  \textbf{Saranya Varakunan}\textsuperscript{1} (ORCID: 0009-0004-0982-9410)\\
  \textbf{Melissa Stadt}\textsuperscript{1} (ORCID: 0000-0002-4215-8004)\\
  \textbf{Mohammad Kohandel}\textsuperscript{1} (ORCID: 0000-0003-0667-7269)
  \par}

  \vspace{1em}
  {\small
  \textsuperscript{1}Department of Applied Mathematics, University of Waterloo,\\
  Waterloo, Ontario, Canada N2L 3G1\par}

  \vspace{4em}
  {\small
  \textbf{Corresponding author:} Saranya Varakunan (saranya.varakunan@uwaterloo.ca)
  \par}

  \vfill
  \thispagestyle{empty}
\end{titlepage}


\begin{abstract}
Chimeric antigen receptor (CAR) T cell therapy has shown remarkable success in hematological malignancies, yet patient responses remain highly variable and the roles of CD4$^{+}$ and CD8$^{+}$ subsets are not fully understood. We present an extended mathematical framework of CAR-T cell dynamics that explicitly models CD4$^{+}$ helper and CD8$^{+}$ cytotoxic lineages and their interactions with tumor antigen burden. Building on the Kirouac et al.\ (2023) model of antigen-regulated memory--effector--exhaustion transitions, our system of differential equations incorporates CD4$^{+}$-mediated modulation of CD8$^{+}$ proliferation, cytotoxicity, and memory regeneration through biologically grounded, saturating interactions. Sensitivity analyses identify effector proliferation, antigen turnover, and CD8$^{+}$ expansion rates as dominant drivers of treatment outcome. Virtual patient simulations recover reported qualitative trends in CAR-T composition, including enhanced expansion and tumor clearance for defined CD4:CD8 products relative to CD8-only formulations, while also revealing inter-patient variability and time-dependent effects. To assess the practical limits of patient-level prediction under parameter uncertainty, we introduce controlled noise into key parameters and show that direct mechanistic classification rapidly degrades. We then demonstrate that a simple feed-forward neural network can partially recover predictive signal from noisy inputs, outperforming a naïve baseline while remaining consistent with mechanistic sensitivities. This work positions the extended model as a hypothesis generator, and illustrates how data-driven methods can complement mechanistic modeling when parameter uncertainty constrains predictive confidence.

\end{abstract}

\section{Introduction}
Cancer is a group of diseases characterized by the uncontrolled proliferation of abnormal cells; it arises when various regulatory mechanisms governing cell growth and division become dysfunctional (Hanahan et al. 2022). While traditional treatments such as surgery, radiation, and chemotherapy remain the backbone of cancer care (and are themselves continually being improved upon), immunotherapy has emerged as a promising treatment avenue. The immune system has been causally linked to cancer surveillance and eradication (Hanahan et al. 2022); as such, immunotherapies seek to leverage the body’s own immune system to target and eliminate cancer cells. In particular, CAR-T cell therapy has shown remarkable success in treating hematological malignancies (Mitra et al. 2023). CAR-T cell therapy involves sampling the patient’s T cells, which are then genetically engineered to express a chimeric antigen receptor (CAR) that allows detection of cancer-specific proteins (antigens). These cancer-targeting cells are expanded in number and reinfused into the patient, where they expand in the body and mount a targeted immune response by recognizing and eliminating cancer cells that display the specific antigen.

Despite being an innovative intersection of genetic engineering and immunology, CAR-T therapy faces significant challenges; these include variable patient responses, potentially life-threatening side effects, and immune suppression (Sterner et al. 2021). These complexities, particularly the unpredictability of patient responses, highlight the need for a deeper understanding of the dynamic interactions between CAR-T cells and cancer cells. Mathematical modeling provides a framework to simulate, analyze, and predict cancer treatment outcomes (Ghaffari et al. 2022). Our model builds upon the mechanistic framework proposed by Kirouac et al. (Kirouac et al. 2023), which describes CAR-T cell pharmacokinetics using antigen-regulated transitions between memory, effector, and exhausted T cell states. The Kirouac et al. model was developed by integrating clinical response data with transcriptomic profiling of CAR-T products, revealing that the cell-state composition of CAR-T therapy was predictive of treatment success. By modeling transitions between memory (T\textsubscript{M}), early effector (T\textsubscript{E1}), cytotoxic effector (T\textsubscript{E2}), and exhausted (T\textsubscript{X}) states, the framework provides a mechanistic basis for understanding patient variability and optimizing CAR-T cell manufacturing. The Kirouac et al. model also accounts for the discrepancy between the administered CAR‑T dose and the observed initial cell counts; a fraction of the infused dose ($f_{\text{loss}}$) is lost immediately upon infusion. The remaining cells are then instantaneously partitioned into the four T‑cell compartments according to dose‐fraction parameters $\mathrm{fraction}_{TM}$, $\mathrm{fraction}_{TE1}$, $\mathrm{fraction}_{TE2}$, and $\mathrm{fraction}_{TX}$.

Memory, effector, and exhausted states refer to the different functional stages of T cells, each playing a critical role in the immune response (Reiser et al. 2016; Kurachi et al. 2019). Memory T cells (T\textsubscript{M}) are long-lived and can rapidly respond to a previously encountered antigen, providing a quick and robust immune response upon re-exposure. Effector T cells (T\textsubscript{E1}, T\textsubscript{E2}) are activated T cells that proliferate and differentiate to perform their primary function of eliminating infected or cancerous cells.  However, during prolonged antigen exposure, such as in chronic infections or cancers, T cells may transition into an exhausted state (T\textsubscript{X}), losing their ability to effectively eliminate target cells. Exhaustion of T cells is a significant barrier in cancer immunotherapy, as it impairs the ability of immune cells to effectively target and eliminate tumor cells.
Note that this transition between states is dynamic, meaning an individual T cell can switch from one state to another depending on a variety of factors (duration and intensity of antigen exposure, signals from surrounding cells, etc). 

CD4\textsuperscript{+} and CD8\textsuperscript{+} T cells are distinct lineages of T cells (Germain et al. 2002), each with unique and complementary roles in the immune response. CD8\textsuperscript{+} T cells, known as "cytotoxic" T cells, specialize in directly killing foreign cells (Andersen et al. 2006). They are equipped with the ability to recognize antigens presented by infected or malignant cells and trigger apoptosis (cell death) in those cells. CD4\textsuperscript{+} T cells, often referred to as "helper" T cells, play a crucial role in orchestrating the immune response by secreting cytokines that activate other immune cells, including CD8\textsuperscript{+} T cells (Ahrends et al., 2013; Ahrends et al., 2017). Importantly, CD4\textsuperscript{+} and CD8\textsuperscript{+} T cells are separate lineages; they do not switch from one to the other, but instead, each lineage maintains its specialized function throughout its lifespan. In cancer immunotherapy, the roles of both CD4\textsuperscript{+} and CD8\textsuperscript{+} T cells are critical, as CD4\textsuperscript{+} cells provide necessary signals to optimize CD8\textsuperscript{+} T cell activity, creating a synergistic effect in tumor elimination (Boulch et al. 2021).

It has been shown that a 1:1 ratio of CD4\textsuperscript{+} to CD8\textsuperscript{+} CAR-T cells yields optimal therapeutic outcomes (as opposed to an unspecified ratio dependent on patient T cell composition), balancing efficacy and toxicity across multiple dosing levels (Turtle et al. 2016). Motivated by these findings, we extend the Kirouac et al. model to explicitly include CD4\textsuperscript{+} and CD8\textsuperscript{+} T cell lineages and their distinct functional roles. This formulation enables the simulation of various CD4:CD8 ratios and their impact on tumor–immune dynamics. We extend this framework by modeling separate CD4\textsuperscript{+} and CD8\textsuperscript{+} CAR-T cell lineages, thereby enabling the investigation of their distinct functional roles and synergistic interactions (Figure \ref{fig:car_t_model}). Importantly, the model does not explicitly impose any optimal CD4:CD8 ratio; rather, we explore how effective tumor clearance may emerge from the complementary requirements of sufficient CD8\textsuperscript{+} cytotoxic effector activity and CD4\textsuperscript{+} helper-mediated support. In particular, the 1:1 ratio is examined because of its clinical relevance and established use in defined-composition CAR-T products.

While these mechanistic extensions capture many aspects of CAR-T cell biology, their predictive use in heterogeneous patient populations can be limited by parameter uncertainty and measurement noise. To address this, we also investigate a feed-forward neural network trained on noisy virtual patient data as a complementary tool for response prediction. To ensure interpretability and maintain consistency with mechanistic understanding, we apply SHAP analysis to compare the network’s learned feature importance with traditional sensitivity analyses.

The goal of this work is therefore twofold. First, we extend the Kirouac et al. framework by explicitly separating CD4\textsuperscript{+} and CD8\textsuperscript{+} CAR-T cell populations in order to mechanistically examine how the ratio of these two lineages (i.e. the balance of helper and cytotoxic roles) may shape treatment outcomes. Second, we assess the extent to which such a mechanistic model can support patient-level outcome prediction in the presence of biological heterogeneity and measurement noise. Rather than performing a comprehensive bifurcation or dynamical systems analysis, our focus is on identifying which parameters most strongly influence clinically relevant outcomes, and evaluating how robust prediction is dependent on measurement of these parameters.

\section{Methods}

\begin{figure}[h!]
  \centering
  \includegraphics[width=0.75\textwidth]{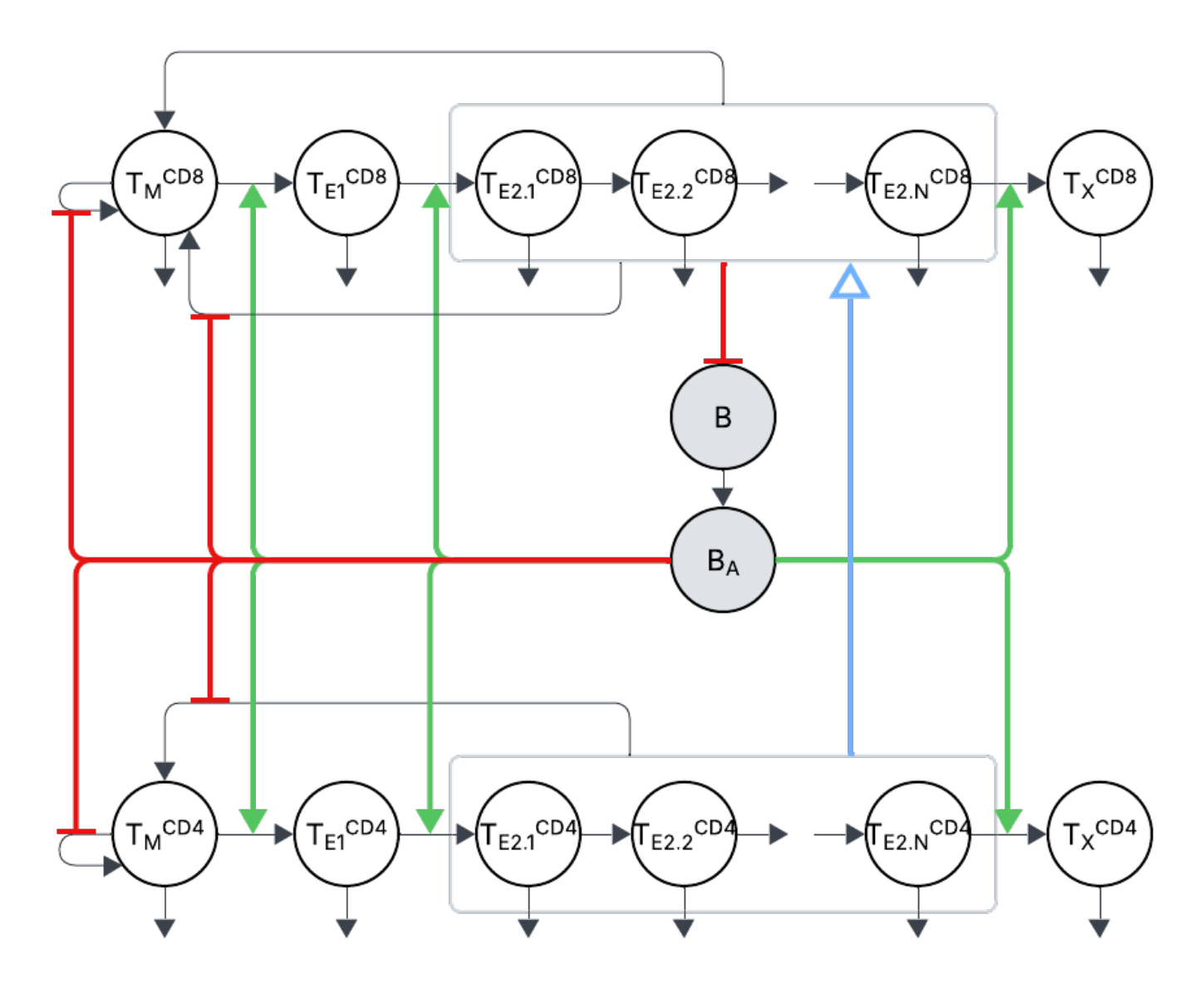}
  \caption{\textbf{Fig. 1} Schematic diagram of our extension of the Kirouac et al. (2023) CAR-T cell model. CD8\textsuperscript{+} T cells (top layer) undergo antigen-driven transitions from memory (T\textsubscript{M}) to effector (T\textsubscript{E1}, T\textsubscript{E2}) to exhausted (T\textsubscript{X}) states, with T\textsubscript{E2}\textsuperscript{CD8} mediating tumor killing. CD4\textsuperscript{+} CAR-T cells (bottom layer) follow a similar differentiation pathway but do not kill the tumor directly. Instead, T\textsubscript{E2}\textsuperscript{CD4} cells exert helper functions by secreting cytokines (e.g., IFN-\(\gamma\), IL-2) which enhance CD8\textsuperscript{+} T cell expansion, survival, and memory formation (blue arrow). Tumor cells (B) generate a transient antigen signal (B\textsubscript{A}) that regulates T cell responses (green and red arrows)}
  \label{fig:car_t_model}
\end{figure}

In this study, we develop a coupled mathematical model as a minimal extension of the CAR-T cell framework proposed by Kirouac et al.\ (2023). Tumor growth, antigen dynamics, and the antigen-regulated memory-effector-exhaustion structure of CAR-T cells are retained unchanged. The primary extension is the explicit separation of the CAR-T population into distinct CD4\textsuperscript{+} and CD8\textsuperscript{+} lineages, together with CD4\textsuperscript{+}-mediated modulation of CD8\textsuperscript{+} cytotoxicity and memory regeneration. When the CD4\textsuperscript{+} dose is set to zero, all CD4\textsuperscript{+} compartments remain identically zero and the CD4-mediated modulation terms reduce to unity, such that the system collapses to the original Kirouac et al.\ formulation (up to relabeling of the aggregate CAR-T population). This structure enables systematic exploration of CD4:CD8 composition effects while preserving the original model as a limiting case. By modeling the interactions between these cell lineages and tumor cells, we aim to predict the therapeutic outcomes of CAR-T cell therapies with varying CD4:CD8 ratios. The model also allows for an exploration of the effects of patient-specific variations on treatment response. A schematic representation of the model is shown in Figure \ref{fig:car_t_model}.

\subsection{Tumor and Antigen Dynamics}
Kirouac et al. developed a mathematical model of cancer based on hematological malignancies, specifically acute lymphoblastic leukemia (ALL) and large B cell lymphoma (LBCL), which are treated with CD19-targeted CAR-T cell therapy (2023). Both ALL and LBCL are B cell-derived tumors, characterized by the uncontrolled proliferation of B cells. CD19 CAR-T therapy targets the CD19 protein, which is predominantly expressed on the surface of B cells, enabling the engineered T cells to selectively eliminate B cells in these cancers. 

The mechanistic framework proposed by Kirouac et al. involves logistic growth of the cancer cell population \(B\), with CAR‑T mediated killing (Eq. (1)). Tumor antigen \(B_A\) is generated in proportion to tumor burden, and cleared linearly (Eq. (2)).

\begin{align}
\small
\frac{dB}{dt} &= \mu_B \left(1 - \frac{B}{B_{\max}} \right) B 
- k_{\text{kill}}^{CD8} \left(1 + \eta_k \, \frac{T_{E2}^{CD4}}{K_\eta + T_{E2}^{CD4}} \right) 
\, \left( \frac{{T_{E2}^{CD8}}^{k_t}}{TK50^{k_t} + {T_{E2}^{CD8}}^{k_t}} \right) B \nonumber \\
&- k_{\text{kill}}^{CD4} \left( \frac{{T_{E2}^{CD4}}^{k_t}}{TK50^{k_t} + {T_{E2}^{CD4}}^{k_t}} \right) B \\[1ex]
\frac{dB_A}{dt} &= k_{B1} B - k_{B2} B_A
\end{align}

In this model, we introduce a mechanistic distinction between CD4\textsuperscript{+} and CD8\textsuperscript{+} CAR-T cells by allowing CD4\textsuperscript{+} cells to  play a major modulatory role in CD8\textsuperscript{+} killing. Specifically, the cytotoxic killing rate of CD8\textsuperscript{+} cells is scaled by a Hill-type function of the CD4\textsuperscript{+} effector population, representing cytokine-mediated helper effects. This modulation is modeled as:
\[
k_{\text{kill}}^{\text{eff}} = k_{\text{kill}}^{CD8} \left(1 + \eta_k \, \frac{T_{E2}^{CD4}}{K_\eta + T_{E2}^{CD4}} \right),
\]
where \( \eta_k \) denotes the maximum fold-increase in CD8\textsuperscript{+} cytotoxicity attributable to CD4\textsuperscript{+} T cell-derived cytokines, and \( K_\eta \) is the half-saturation constant (Ahrends et al. 2017). Biologically, this formulation reflects the role of CD4\textsuperscript{+} T cells, particularly Th1-like CAR-T cells, in secreting cytokines such as interferon-gamma (IFN-\(\gamma\)) and interleukin-2 (IL-2), which are known to enhance the activation, survival, and cytotoxic function of CD8\textsuperscript{+} T cells (Boulch et al. 2021). IFN-\(\gamma\) also promotes tumor antigen presentation and induces tumor cell apoptosis; accordingly, the final term in Eq.~(1) represents an effective CD4\textsuperscript{+}-mediated contribution to tumor control through cytokine signaling rather than direct cytotoxic killing (Bove et al., 2023). 

CD4\textsuperscript{+} T cells are also known to increase the regeneration of memory CD8\textsuperscript{+} T cells from effector CD8\textsuperscript{+} T cells. As such, a Hill-type function of the CD4\textsuperscript{+} effector cells was also included as a factor in the regeneration term for CD8\textsuperscript{+} memory T cells, with parameter estimates from relevant literature (Ahrends et al, 2013).

\subsection{CD8\textsuperscript{+} CAR-T Cell Dynamics}

CD8\textsuperscript{+} T cells, also known as cytotoxic T cells, are critical for directly eliminating infected or cancerous cells (Andersen et al. 2006). They recognize tumor antigens presented by infected or malignant cells and are activated; then, they replicate and differentiate into effector cells that exert cytotoxic effects, including apoptosis (induced cell death) in tumor cells. The function of CD8\textsuperscript{+} T cells in CAR-T therapy is to recognize and kill tumor cells expressing specific antigens, such as CD19, on their surface. Their effectiveness can be influenced by various factors, including antigen availability, T cell exhaustion, and modulation by CD4\textsuperscript{+} T cells through cytokine secretion.

The dynamics of CD8\textsuperscript{+} CAR-T cells are modeled by a set of differential equations that model the transitions between different T cell states: memory ($T_{M}^{CD8}$), early effector ($T_{E1}^{CD8}$), late effector ($T_{E2}^{CD8}$), and exhausted ($T_{X}^{CD8}$) (Kirouac et al. 2023). These equations take into account the proliferation, differentiation, death rates, and interactions with tumor antigens (\(B_A\)).

The memory CD8\textsuperscript{+} T cells (\(T_M^{CD8}\)) proliferate based on antigen recognition and a maximum self-renewal rate (\(f_{\text{max}}^{CD8}\)), with modulation from CD4\textsuperscript{+} effector T cells (\(T_{E2}^{CD4}\)) (4). Effector CD8\textsuperscript{+} T cells differentiate from memory cells (\(T_M^{CD8}\)) and are affected by antigen availability, tumor burden, and CD4-mediated cytokine modulation (5). The late effector T cells (\(T_{E2}^{CD8}\)) mediate direct tumor killing (6), while the exhaustion state (\(T_X^{CD8}\)) models the decline in cell function due to prolonged antigen exposure (7).

These equations incorporate Hill-type functions to model the saturation effects of antigen on T cell proliferation and exhaustion, with parameters such as \(B_{50}\), \(k_m^{CD8}\), and \(k_e^{CD8}\) controlling these relationships. Additionally, parameters such as \(f_{\text{Tm}}^{CD8}\) and \(f_{\text{Te2}}^{CD8}\) govern the T cell expansion and response to treatment, reflecting the impact of the administered CAR-T cell dose. 

\begin{align}
\small 
\frac{dT_M^{CD8}}{dt} &= \mu_M^{CD8} \left( 2 f_{\text{max}}^{CD8} \left(1 - \frac{B_A^{k_m^{CD8}}}{B50^{CD8^{k_m}} + B_A^{k_m^{CD8}}} \right) - 1 \right) T_M^{CD8} \nonumber \\
&\quad + r_M^{CD8} \left(1 + \eta_r \, \frac{T_{E2}^{CD4}}{K_\eta + T_{E2}^{CD4}} \right) \left(1 - \frac{B_A^{k_r^{CD8}}}{B50^{CD8^{k_r}} + B_A^{k_r^{CD8}}} \right) T_{E2}^{CD8} \nonumber \\
&\quad - d_M^{CD8} T_M^{CD8} + f_{\text{Tm}}^{CD8} \cdot \text{Dose}_{CD8} \\[1ex]
\frac{dT_{E1}^{CD8}}{dt} &= 2 \mu_M^{CD8} \left(1 - f_{\text{max}}^{CD8} \left(1 - \frac{B_A^{k_m^{CD8}}}{B50^{CD8^{k_m}} + B_A^{k_m^{CD8}}} \right) \right) T_M^{CD8} \nonumber \\
&\quad - \mu_E^{CD8} \left( \frac{B_A^{k_e^{CD8}}}{B50^{CD8^{k_e}} + B_A^{k_e^{CD8}}} \right) T_{E1}^{CD8} - d_{E1}^{CD8} T_{E1}^{CD8} + f_{\text{Te1}}^{CD8} \cdot \text{Dose}_{CD8} \\[1ex]
\frac{dT_{E2}^{CD8}}{dt} &= \mu_E^{CD8} \cdot 2^N \cdot \left( \frac{B_A^{k_e^{CD8}}}{B50^{CD8^{k_e}} + B_A^{k_e^{CD8}}} \right) T_{E1}^{CD8} \nonumber \\
&\quad - k_{ex}^{CD8} \left( \frac{B_A^{k_x^{CD8}}}{B50^{CD8^{k_x}} + B_A^{k_x^{CD8}}} \right) T_{E2}^{CD8} \nonumber \\
&\quad - r_M^{CD8} \left(1 + \eta_r \, \frac{T_{E2}^{CD4}}{K_\eta + T_{E2}^{CD4}} \right) \left(1 - \frac{B_A^{k_r^{CD8}}}{B50^{CD8^{k_r}} + B_A^{k_r^{CD8}}} \right) T_{E2}^{CD8} \nonumber \\
&\quad - d_{E2}^{CD8} T_{E2}^{CD8} + f_{\text{Te2}}^{CD8} \cdot \text{Dose}_{CD8} \\[1ex]
\frac{dT_X^{CD8}}{dt} &= k_{ex}^{CD8} \left( \frac{B_A^{k_x^{CD8}}}{B50^{CD8^{k_x}} + B_A^{k_x^{CD8}}} \right) T_{E2}^{CD8} - d_X^{CD8} T_X^{CD8} + f_{\text{Tx}}^{CD8} \cdot \text{Dose}_{CD8}
\end{align}

\subsection{CD4\textsuperscript{+} CAR-T Cell Dynamics}

CD4\textsuperscript{+} T cells, or helper T cells, play an essential role in coordinating the immune response. Unlike CD8\textsuperscript{+} T cells, the main role of CD4\textsuperscript{+} cells is not directly killing infected or cancerous cells (although they can induce cell death by themselves) (Boulch et al. 2021). Instead, their main function involves enhancing the function of other immune cells, including CD8\textsuperscript{+} T cells, through the secretion of cytokines such as interferon-gamma (IFN-\(\gamma\)) and interleukin-2 (IL-2) (Ahrends et al. 2017). These cytokines stimulate the expansion and survival of CD8\textsuperscript{+} T cells, thereby augmenting their cytotoxic effects on tumor cells. In CAR-T therapy, CD4\textsuperscript{+} T cells are critical for sustaining an effective immune response, promoting memory formation, and improving tumor clearance (Boulch et al. 2021). CD4\textsuperscript{+} modulation of CD8\textsuperscript{+} function is represented using a saturating Hill-type term, reflecting cytokine-mediated helper effects that are nonlinear and saturable in nature. This choice provides a minimal and biologically grounded interaction without introducing additional state variables or exacerbating parameter identifiability issues.

The equations governing the dynamics of CD4\textsuperscript{+} T cells are analogous to those of CD8+ T cells, with the exception of the modulation of CD8\textsuperscript{+} memory cell renewal by CD4\textsuperscript{+} T cells through cytokine-mediated interactions.

\begin{align}
\small
\frac{dT_M^{CD4}}{dt} &= \mu_M^{CD4} \left( 2 f_{\text{max}}^{CD4} \left(1 - \frac{B_A^{k_m^{CD4}}}{B50^{CD4^{k_m}} + B_A^{k_m^{CD4}}} \right) - 1 \right) T_M^{CD4} \nonumber \\
&\quad + r_M^{CD4} \left(1 - \frac{B_A^{k_r^{CD4}}}{B50^{CD4^{k_r}} + B_A^{k_r^{CD4}}} \right) T_{E2}^{CD4} - d_M^{CD4} T_M^{CD4} + f_{\text{Tm}}^{CD4} \cdot \text{Dose}_{CD4} \\[1ex]
\frac{dT_{E1}^{CD4}}{dt} &= 2 \mu_M^{CD4} \left(1 - f_{\text{max}}^{CD4} \left(1 - \frac{B_A^{k_m^{CD4}}}{B50^{CD4^{k_m}} + B_A^{k_m^{CD4}}} \right) \right) T_M^{CD4} \nonumber \\
&\quad - \mu_E^{CD4} \left( \frac{B_A^{k_e^{CD4}}}{B50^{CD4^{k_e}} + B_A^{k_e^{CD4}}} \right) T_{E1}^{CD4} - d_{E1}^{CD4} T_{E1}^{CD4} + f_{\text{Te1}}^{CD4} \cdot \text{Dose}_{CD4} \\[1ex]
\frac{dT_{E2}^{CD4}}{dt} &= \mu_E^{CD4} \cdot 2^N \cdot \left( \frac{B_A^{k_e^{CD4}}}{B50^{CD4^{k_e}} + B_A^{k_e^{CD4}}} \right) T_{E1}^{CD4} \nonumber \\
&\quad - k_{ex}^{CD4} \left( \frac{B_A^{k_x^{CD4}}}{B50^{CD4^{k_x}} + B_A^{k_x^{CD4}}} \right) T_{E2}^{CD4} \nonumber \\
&\quad - r_M^{CD4} \left(1 - \frac{B_A^{k_r^{CD4}}}{B50^{CD4^{k_r}} + B_A^{k_r^{CD4}}} \right) T_{E2}^{CD4} - d_{E2}^{CD4} T_{E2}^{CD4} + f_{\text{Te2}}^{CD4} \cdot \text{Dose}_{CD4} \\[1ex]
\frac{dT_X^{CD4}}{dt} &= k_{ex}^{CD4} \left( \frac{B_A^{k_x^{CD4}}}{B50^{CD4^{k_x}} + B_A^{k_x^{CD4}}} \right) T_{E2}^{CD4} - d_X^{CD4} T_X^{CD4} + f_{\text{Tx}}^{CD4} \cdot \text{Dose}_{CD4}
\end{align}

\subsection{Model Parameters}

A list of parameters used in the model is given in Table \ref{tab:model_parameters}. Kirouac et al. reported physiologically feasible ranges and averages for the parameters in their original model, which did not have separate CD4\textsuperscript{+} and CD8\textsuperscript{+} compartments (Kirouac et al. 2023). However, several parameters are likely to differ biologically between these two lineages (Table \ref{tab:cd4cd8_ratios}). 

\begin{table}[H]
\centering
\scriptsize
\renewcommand{\arraystretch}{0.9} 
\setlength{\tabcolsep}{4pt}       
\begin{tabular}{|p{0.2\linewidth}|p{0.4\linewidth}|p{0.15\linewidth}|p{0.15\linewidth}|}
\hline
\textbf{Parameter} & \textbf{Description} & \textbf{Value} & \textbf{Source} \\
\hline
$\mu_B$ & Tumor cell proliferation rate & [0.001, 0.1] & \cite{Kirouac2023} \\
$B_{\max}$ & Tumor carrying capacity & [\(10^8\), \(10^{12}\)] & \cite{Kirouac2023} \\
$k_{\mathrm{kill}}^{CD8}$ & CD8-mediated tumor killing rate & [0.001, 1] & \cite{Kirouac2023} \\
$\eta_k$ & CD4-mediated enhancement of CD8 cytotoxicity & 10 & \cite{Ahrends2017} \\
$\eta_r$ & CD4-mediated enhancement of CD8 regeneration to memory cells & 10 & \cite{Ahrends2013} \\
$K_\eta$ & Half-saturation constant for CD4 modulation & \(10^5\) & \cite{Ahrends2017} \\
$TK50$ & Half-max for tumor killing by $T_{E2}^{CD8}$ & [\(10^5\), \(10^9\)] & \cite{Kirouac2023} \\
$k_{B1}$ & Tumor antigen generation rate & [0.001, 1] & \cite{Kirouac2023} \\
$k_{B2}$ & Tumor antigen clearance rate & [0.001, 1] & \cite{Kirouac2023} \\
$\mu_M^{CD4}$, $\mu_M^{CD8}$ & Memory T cell proliferation rate & [0.001, 1] & \cite{Kirouac2023} \\
$d_M^{CD4}$, $d_M^{CD8}$ & Memory T cell death rate & [0.001, 1] & \cite{Kirouac2023} \\
$f_{\mathrm{max}}^{CD4}$, $f_{\mathrm{max}}^{CD8}$ & Max self-renewal fraction of memory T cells & [0.5, 0.99] & \cite{Kirouac2023} \\
$\beta_M^{CD4}$, $\beta_M^{CD8}$ & Hill exponent for memory differentiation & [0.2, 3] & \cite{Kirouac2023} \\
$B50_M^{CD4}$, $B50_M^{CD8}$ & Antigen half-max for memory differentiation & [\(10^6\), \(10^{10}\)] & \cite{Kirouac2023} \\
$\mu_E^{CD4}$, $\mu_E^{CD8}$ & Effector T cell proliferation rate & [0.001, 1] & \cite{Kirouac2023} \\
$d_{E1}^{CD4}$, $d_{E1}^{CD8}$ & Early effector T cell death rate & [0.001, 1] & \cite{Kirouac2023} \\
$\beta_E^{CD4}$, $\beta_E^{CD8}$ & Hill exponent for effector expansion & [0.2, 3] & \cite{Kirouac2023} \\
$B50_E^{CD4}$, $B50_E^{CD8}$ & Antigen half-max for effector expansion & [\(10^6\), \(10^{10}\)] & \cite{Kirouac2023} \\
$k_{ex}^{CD4}$, $k_{ex}^{CD8}$ & Exhaustion induction rate & [0.001, 1] & \cite{Kirouac2023} \\
$d_{E2}^{CD4}$, $d_{E2}^{CD8}$ & Late effector T cell death rate & [0.001, 1] & \cite{Kirouac2023} \\
$d_X^{CD4}$, $d_X^{CD8}$ & Exhausted T cell death rate & [0.001, 1] & \cite{Kirouac2023} \\
$\beta_{ex}^{CD4}$, $\beta_{ex}^{CD8}$ & Hill exponent for exhaustion & [0.2, 3] & \cite{Kirouac2023} \\
$B50_{ex}^{CD4}$, $B50_{ex}^{CD8}$ & Antigen half-max for exhaustion & [\(10^6\), \(10^{10}\)] & \cite{Kirouac2023} \\
$r_M^{CD4}$, $r_M^{CD8}$ & Rate of reversion from effector to memory & [0.001, 1] & \cite{Kirouac2023} \\
$\beta_r^{CD4}$, $\beta_r^{CD8}$ & Hill exponent for reversion & [0.2, 3] & \cite{Kirouac2023} \\
$B50_r^{CD4}$, $B50_r^{CD8}$ & Antigen half-max for reversion & [\(10^6\), \(10^{10}\)] & \cite{Kirouac2023} \\
$k_t$ & B cell killing Hill exponent & [0.2, 3] & \cite{Kirouac2023} \\
$N$ & Number of population doublings in TE2 & [4, 12] & \cite{Kirouac2023} \\
$\text{fraction}_{TM}$ & T memory fraction of dose & [1, 10] & \cite{Kirouac2023} \\
$\text{fraction}_{TE1}$ & T effector (TE1) fraction of dose & [1, 10] & \cite{Kirouac2023} \\
$\text{fraction}_{TE2}$ & T effector (TE2) fraction of dose & [30, 70] & \cite{Kirouac2023} \\
$\text{fraction}_{TX}$ & T exhausted fraction of dose & [5, 30] & \cite{Kirouac2023} \\
\hline
\end{tabular}
\caption{Model parameters used in the extended CD4\textsuperscript{+}/CD8\textsuperscript{+} CAR-T cell model, with value/range and source. Note that for parameters with CD4\textsuperscript{+} and CD8\textsuperscript{+} values, the value reported is the CD8\textsuperscript{+} parameter value. See Table \ref{tab:cd4cd8_ratios} for CD4\textsuperscript{+} parameter values.}
\label{tab:model_parameters}
\end{table}

CD8\textsuperscript{+} CAR-T cells are primarily responsible for direct cytotoxicity and tend to differentiate more rapidly into effector cells (\( \mu_E \)) but are also more prone to exhaustion (\( k_{ex} \)) and activation-induced cell death (\( d_E \), \( d_X \)) (Turtle et al. 2016; Hinrichs et al. 2011; Foulds et al. 2002; Fraietta et al. 2018; Gauthier et al. 2021; Blank et al. 2019; Siddiqui et al. 2019). In contrast, CD4\textsuperscript{+} CAR-T cells play a predominantly helper role through cytokine production (e.g., IFN-\(\gamma\), IL-2), are generally more resistant to exhaustion, and may exhibit greater longevity and plasticity, potentially resulting in higher rates of reversion to memory (\( r_M \)). As such, when fitting the model to data or simulating lineage-specific behavior, these parameters should be relaxed and estimated separately for CD4\textsuperscript{+} and CD8\textsuperscript{+} populations. This refinement is supported by experimental evidence from preclinical and clinical studies, including those by Turtle et al.~(2016), Fraietta et al.~(2018), and Gauthier et al.~(2021), which consistently demonstrate phenotypic and functional distinctions between CD4\textsuperscript{+} and CD8\textsuperscript{+} CAR-T cells in vivo. 

\begin{table}[H]
\centering
\scriptsize
\renewcommand{\arraystretch}{1.4}
\begin{tabular}{|l|p{4.5cm}|c|p{6.5cm}|}
\hline
\textbf{Parameter} & \textbf{Definition} & \textbf{\(X_{CD4}/X_{CD8}\)} & \textbf{Rationale} \\
\hline
\( \mu_E \) & Effector differentiation rate & \(\sim 0.6\text{–}0.8\) & CD4\textsuperscript{+} T cells generally differentiate more slowly into effector cells than CD8\textsuperscript{+} cells, with delayed peak proliferation and cytokine secretion kinetics~\cite{Turtle2016, Hinrichs2011, Foulds2002, Fraietta2018}. \\
\hline
\( d_E, d_X \) & Effector T cell, exhausted T cell death rates & \(\sim 0.5\text{–}0.7\) & CD4\textsuperscript{+} T cells are more resistant to activation-induced cell death and may survive longer than CD8\textsuperscript{+} cells in chronic stimulation environments~\cite{Fraietta2018, Gauthier2021}. \\
\hline
\( k_{ex} \) & Exhaustion rate & \(\sim 0.1\text{–}0.2\) & CD8\textsuperscript{+} T cells show higher PD-1, TOX, and TIM-3 expression under chronic antigen load, and are more prone to exhaustion than CD4\textsuperscript{+} T cells~\cite{Blank2019}. \\
\hline
\( r_M \) & Differentiation from effector to memory T cell & \(\sim 1.2\text{–}1.5\) & CD4\textsuperscript{+} T cells exhibit more plasticity and longevity, with a higher tendency to persist and transition into memory phenotypes~\cite{Siddiqui2019}. \\
\hline
\( \mu_M, d_M \) & Memory T cell proliferation rate, death rate & \(\sim 1.0\) & Memory proliferation rates are generally comparable for CD4\textsuperscript{+} and CD8\textsuperscript{+} T cells, although long-term persistence may vary depending on subset composition~\cite{Gattinoni2011}. \\
\hline
\( k_{\text{kill}} \) & Per-cell tumor killing rate & \(\sim 0.7\text{–}0.9\) &  
CD4\textsuperscript{+} CD4 CAR-T cells indirectly cause tumor cell death (e.g.,  
IFN-\(\gamma\), TNF-$\alpha$–mediated macrophage activation) and do not use direct cytotoxicity like CD8\textsuperscript{+} cells, resulting in a slightly reduced per-cell kill efficiency~\cite{Bove2023}. \\[-0.3em]
\hline
\end{tabular}
\caption{Suggested parameter ratios between CD4\textsuperscript{+} and CD8\textsuperscript{+} CAR-T cells based on biological considerations and literature reports. All other parameters are assumed to be identical between CD4 and CD8 lineage cells.}
\label{tab:cd4cd8_ratios}
\end{table}

\subsection{Sensitivity Analysis}
Sensitivity analysis methods are used to determine how variations in model parameters impact model output (Qian et al. 2020). This method may be used to identify which parameters have the most significant impact on the system's behavior, thereby guiding model refinement, parameter selection, and interpretation of results. Additionally, it allows us to understand the robustness of model predictions and the effects of parameter uncertainty, providing valuable insights for optimizing treatment strategies and predicting patient responses. 

Because the Kirouac et al. framework is parameterized using high-dimensional, correlated distributions derived from patient data, isolating low-dimensional bifurcation structure is not straightforward. While nonlinear transitions in treatment outcome can be observed when varying individual parameters or parameter pairs, these thresholds are not fixed across patients and are strongly modulated by variability in other biological processes. For this reason, we focus on sensitivity-based analyses that better reflect clinically relevant heterogeneity.
 
There are two main types of sensitivity analysis: local and global. Local sensitivity analysis involves perturbing each parameter individually from a baseline value while holding other parameters constant, and looking at the change in output (Qian et al. 2020). In this context, we applied finite difference perturbations (1\%) to each parameter (starting from baseline patient parameter values) and initial condition individually while tracking changes in the area under the curve (AUC) of tumor burden over 1 year. Sensitivities were computed as normalized relative changes, defined as the relative change in the output (area under the curve) normalized by the relative change in the input, allowing us to rank the most influential parameters under the baseline scenario.

While local sensitivity analysis looks at the effect of changing one parameter at a time, global sensitivity analysis assesses the impact of varying multiple parameters simultaneously over the full range of possible values (Qian et al. 2020). Since this approach takes into account the interactions between parameters, it is more reflective of heterogeneity observed in patients. We performed global sensitivity analyses on the tumor-immune ODE model using both Partial Rank Correlation Coefficients (PRCC) and Sobol sensitivity analysis, with the same output of interest as in the local sensitivity analysis (AUC of tumor burden (\(B\)) over 1 year). PRCC identifies the strength and direction of monotonic relationships between parameters and outputs, while Sobol analysis quantifies how much each parameter (and combinations of parameters) contributes to the overall variability in the output, such as tumor burden over a year.

 For the PRCC analysis, we applied Latin Hypercube Sampling (LHS) to systematically vary each model parameter within \(\pm 20\%\) of its baseline value across 300 samples, computing PRCC values between the ranked parameter values and ranked AUC outcomes to quantify the strength and direction of monotonic relationships while controlling for the effects of other parameters. For the Sobol analysis, we used Sobol sequence sampling to generate 1000 parameter sets within the same \(\pm 20\%\) ranges and computed first-order (\(S_i\)) and total-order (\(S_{Ti}\)) Sobol indices, quantifying the proportion of variance in tumor burden attributable to each parameter alone and in interaction with others. 

\subsection{Virtual Patient Generation}

Patient data for model validation is often challenging to obtain due to limitations such as a lack of patients, ethical guidelines, and difficulties in obtaining the required measurements. As a result, 'virtual patients' are commonly used in mathematical modeling, especially in biological contexts (Craig et al. 2023). A virtual patient is a synthetic representation of a real patient, defined by a set of parameters that can be input into the mathematical model and simulated to predict outcomes, such as tumor burden. Generating virtual patients involves selecting appropriate parameter values from predefined ranges, typically derived from clinical studies or literature. In this study, we employ random sampling of parameter values within the ranges provided by Kirouac et al. (2023), enabling the simulation of a large and diverse set of virtual patients to explore various treatment scenarios and predict tumor progression.

\subsection{Feed-forward Neural Networks}

Mathematical models provide a framework for identifying key parameters and understanding dynamic relationships. Ultimately, it would be useful for the model to help stratify patients and predict optimal treatments. However, measuring these parameters introduces noise, which can limit the direct applicability of the model. Machine learning methods can complement the model by learning robust patterns from noisy inputs, improving predictive accuracy.
In particular, neural networks are computational models inspired by the structure and function of the brain, capable of learning complex relationships between input features and outputs (Sazli et al. 2006). They are commonly used in biomedical applications for tasks such as classifying patients based on predicted treatment outcomes (Falavigna et al, 2019). A feed-forward neural network (FNN) is a type of neural network in which information flows in one direction, from the input layer through one or more hidden layers to the output layer (Sazli et al. 2006). Each layer consists of neurons, which are processing units that combine weighted inputs and apply an activation function to produce an output. Hidden layers allow the network to capture nonlinear relationships between parameters, while the output layer converts the final computation into a prediction. We implemented a FNN using virtual patient results with noisy input parameters; this would allow the model to learn robust patterns from the simulation results, improving its ability to generalize to new virtual patients even if there is noise in parameter measurements.

\subsection{SHAP Analysis}

To further interpret the predictions made by the neural network, we applied SHAP (SHapley Additive exPlanations) values, a tool for explaining predictions of machine learning models (Lundberg et al. 2017). SHAP quantifies the contribution of each input feature to the network’s output, enabling us to understand which parameters the network considers most influential for predicting treatment outcomes. By using SHAP, we can assess parameter importance in the neural network while capturing nonlinear and interactive effects that may not be fully reflected by traditional sensitivity measures such as local perturbations, PRCC, or Sobol analyses. Comparing SHAP rankings with these established sensitivity analyses offers a more nuanced view of how variability in model parameters drives predicted treatment outcomes under uncertainty.

\subsection{Code and Software}
Simulations, virtual patient generation, and sensitivity analyses were implemented in MATLAB. The neural network was implemented in Python using the PyTorch library. 

\section{Results}

\subsection{Sensitivity Analysis}

\begin{figure}[H]
    \centering
    \includegraphics[width=0.7\textwidth]{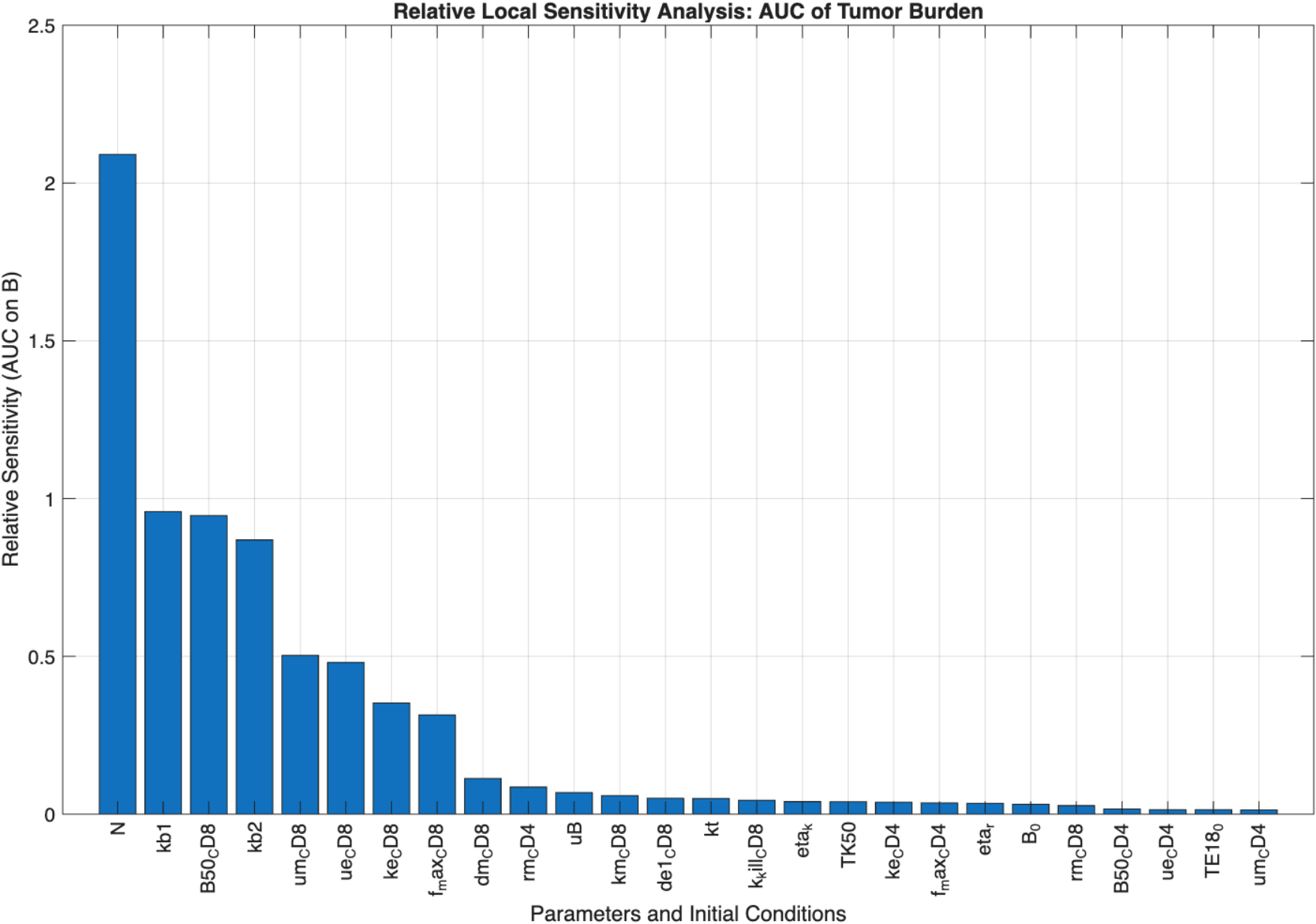} 
    \caption{\textbf{Fig. 2} Relative local sensitivity analysis using the area under the curve (AUC) of tumor burden over 365 days. Bars represent the relative sensitivity of the AUC of \(B(t)\) to a 1\% perturbation in each parameter and initial condition in the model, with parameters with the greatest sensitivities displayed}
    \label{fig:local_sens}
\end{figure}

Local sensitivity analysis (Figure \ref{fig:local_sens}) revealed that the most sensitive parameters were the proliferation burst factor during antigen stimulation (\(N\)), tumor antigen generation and clearance rates (\(k_{b1}\) and \(k_{b2}\)), CD8\textsuperscript{+} T cell expansion dynamics (including the effector proliferation rate \(k_{e}^{CD8}\) and proliferation rate \(\mu_M^{CD8}\)), and tumor-related parameters such as the antigen half-max (\(B_{50}^{CD8}\)) (Figure \ref{fig:local_sens}). These parameters exhibited the highest relative sensitivities; that is, small changes in their values can lead to substantial differences in tumor burden dynamics within the model.

These results agree with the current mechanistic understanding of CAR-T cell therapy, in that expansion of the CAR-T cells (which is dependent on T cell replication rates as well as antigen availability) is key in treatment efficacy. This suggests that the model accurately captures the drivers of tumor-immune dynamics, and is not overly dependent on the separation of the CD4\textsuperscript{+} and CD8\textsuperscript{+} compartments. This supports the model's relevance in simulating and predicting treatment outcomes.

\begin{figure}[H]
    \centering
    \includegraphics[width=0.7\textwidth]{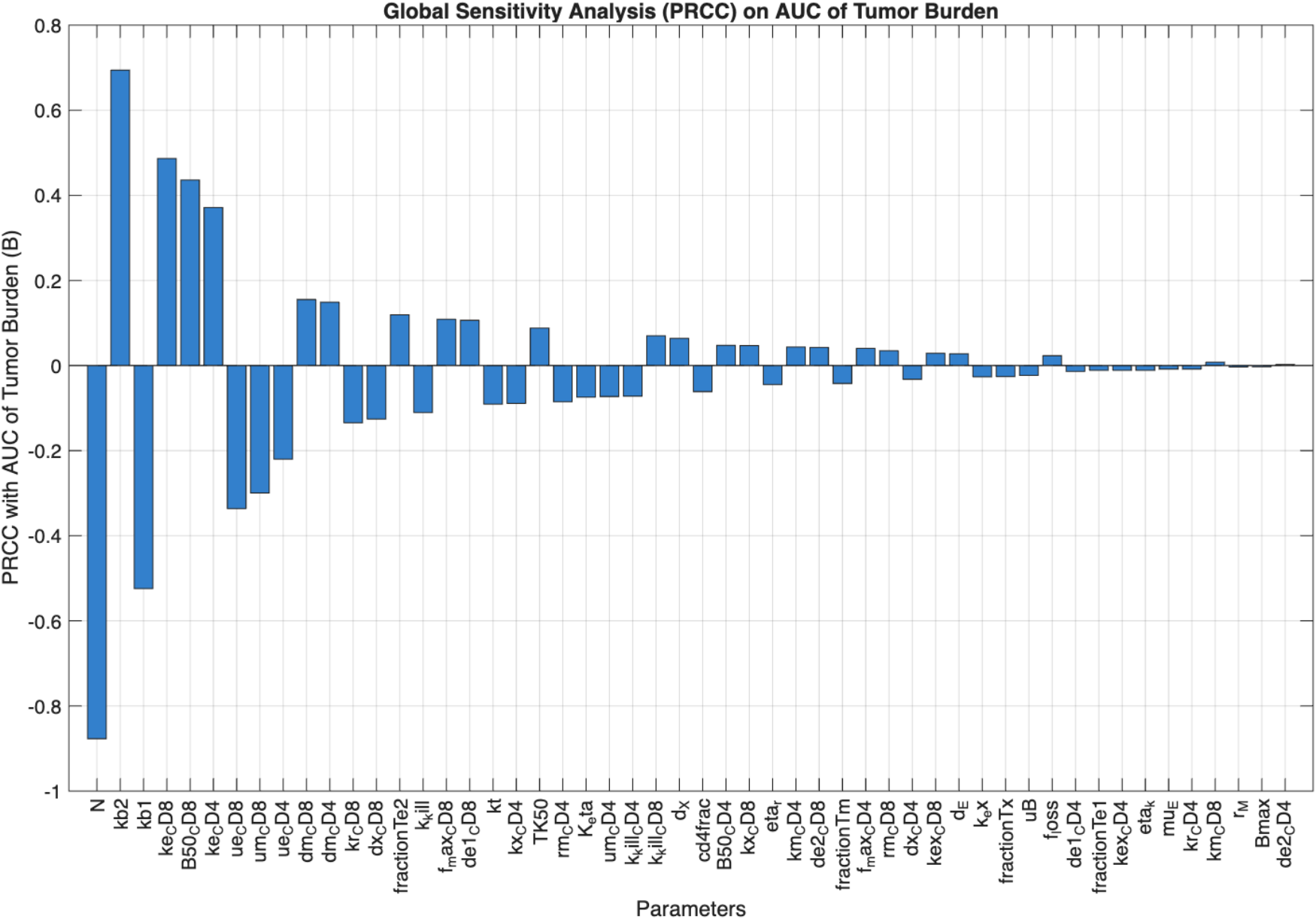} 
    \caption{\textbf{Fig. 3} Global sensitivity analysis using the area under the curve (AUC) of tumor burden over 365 days. Bars represent the partial rank correlation coefficients (PRCC) between each parameter and the AUC of tumor burden, indicating the direction and strength of monotonic relationships while controlling for other parameters. Parameters with the greatest sensitivities are displayed}
    \label{fig:global_sens_prcc}
\end{figure}

\begin{figure}[H]
    \centering
    \includegraphics[width=0.7\textwidth]{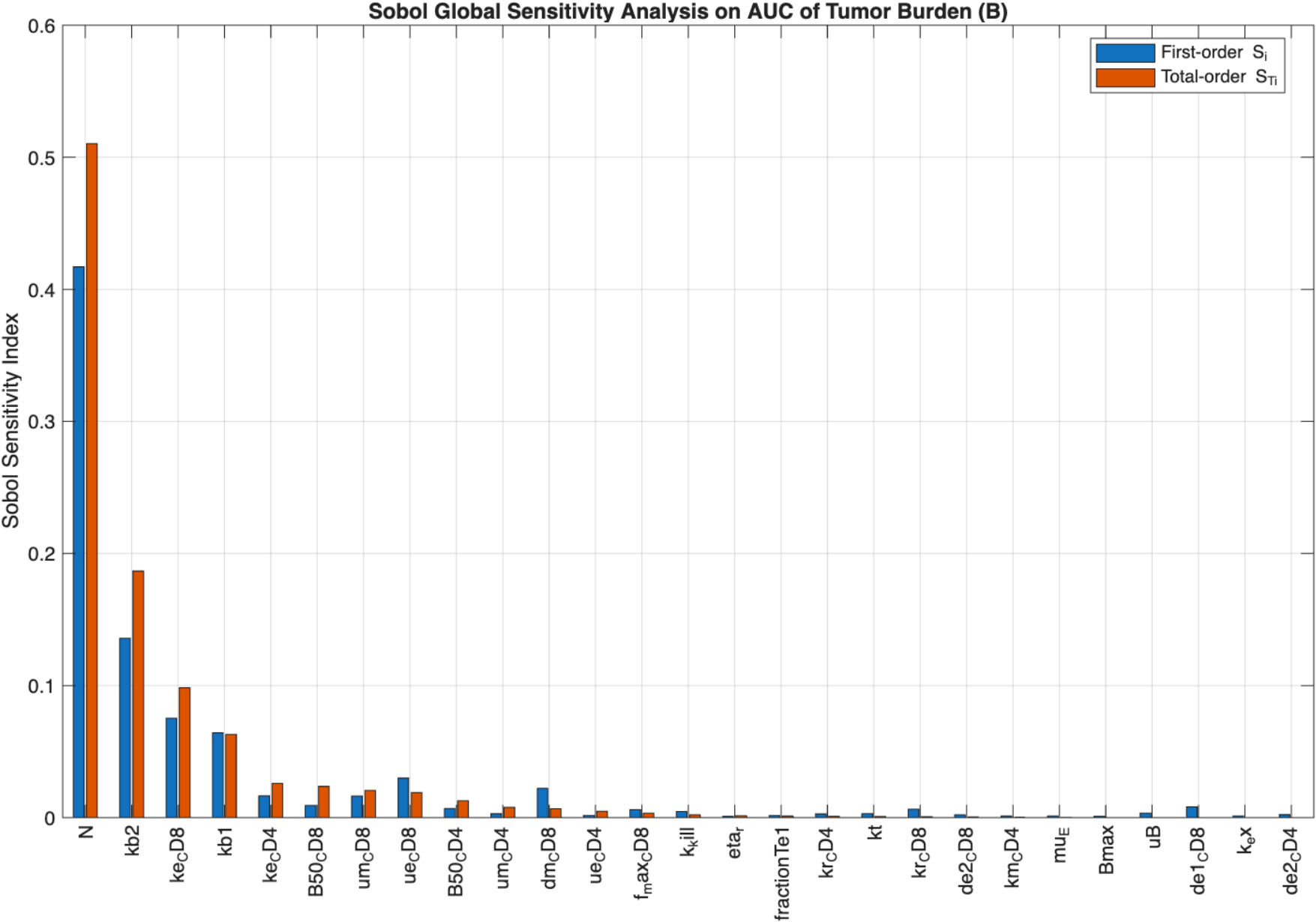} 
    \caption{\textbf{Fig. 4} Sobol global sensitivity analysis using the area under the curve (AUC) of tumor burden over 365 days. Bars represent the first-order ($S_i$) and total-order ($S_{Ti}$) Sobol sensitivity indices for each parameter, quantifying the proportion of variance in tumor burden attributable to each parameter alone and in interaction with others. Only parameters with the greatest sensitivities are displayed}
    \label{fig:global_sens_sobol}
\end{figure}

Global sensitivity analysis (Figures \ref{fig:global_sens_prcc}, \ref{fig:global_sens_sobol}) found that the parameter representing the number of effector cell doublings (\(N\)) exhibits the highest sensitivity in our model, with both PRCC and Sobol analyses indicating it as the strongest driver of tumor burden dynamics. Parameters governing the dynamics of the antigen-bound complex (\(k_{b1}\), \(k_{b2}\)) and the CD8\textsuperscript{+} effector proliferation rate (\(k_{e}^{CD8}\)) also showed high sensitivities. These findings align with prior analysis by Kirouac et al. of their original mechanistic framework (2023), which identified effector proliferation and memory regeneration as critical determinants of treatment response. Our analyses suggest that modulating parameters such as \(N\), \(k_{b2}\), and \(k_{e}^{CD8}\) may have the greatest potential to improve complete response rates in CAR-T therapy by enhancing effector cell expansion and sustaining immune-mediated tumor control under parameter uncertainty.

\subsection{Qualitative consistency of model with reported CAR-T composition effects}

Because suitable clinical datasets containing matched CD4:CD8 ratios, longitudinal CAR-T expansion, and outcome measures remain limited, we do not attempt a full quantitative validation of predicted response or survival. The original Kirouac et al.\ model has been previously analyzed against available clinical data; our extension builds on this framework by explicitly resolving CD4$^+$ and CD8$^+$ CAR-T cell lineages. Accordingly, this section focuses on whether reported qualitative trends in CD4:CD8 composition and treatment outcome can emerge from the extended model under biologically grounded assumptions, positioning the framework as a mechanistic hypothesis generator rather than a finalized clinical prediction tool.

A set of 7500 virtual patients was generated using patient parameter values given in Kirouac et al. (2023); parameters were randomly chosen in the ranges given by patient data. Each patient was simulated for 1000 days under three possible scenarios: (1) no treatment, (2) treatment with CD4:CD8 ratio of 1:1, and (3) treatment with unspecified CD4:CD8 ratio (empty CD4\textsuperscript{+} compartment, as in the original Kirouac et al. model). The no-treatment case serves as a control, the unspecified CD4:CD8 case corresponds to the original Kirouac et al. model with an empty CD4\textsuperscript{+} compartment, and the 1:1 CD4:CD8 treatment is included because of its clinical relevance, as previously described. In each case, patients were categorized into non-responder (NR) or complete responder (CR) depending on the tumor burden (greater or less than $10^9$ cells, respectively) at the end of the simulation time. 

\subsubsection{Consistency of model outcomes with reported effects of 1:1 CAR-T composition}

It has been established by Lee et al that in vivo expansion of CAR-T cells with a 1:1 CD4:CD8 ratio is higher than that of CAR-T cells that only contain CD8 cells (Lee et al. 2018). Since CAR-T expansion is the key factor in tumor elimination, we sought to validate our model using the proportions of virtual patients that responded in each treatment scenario (1:1 treatment versus pure CD8 treatment) as a proxy measurement for CAR-T expansion. Among the virtual patients, it was found that the vast majority (45.03\%, 50.36\%) of patients did not show any change with treatment (Always CR or Always NR); this is to be expected since few real patients respond to CAR-T therapy, and virtual patients who eliminate the tumour without treatment will also be able to eliminate the tumour with treatment. The remaining patients (4.61 \% of virtual patients) showed an improvement of NR to CR with some treatment; 71.2\% of these patients responded to either treatment, 8.73\% responded to CD8 treatment only, and 20.11\% responded to 1:1 treatment only. The higher proportion of patients that responded to 1:1 treatment compared to pure CD8 treatment is in agreement with the aforementioned in vivo expansion data from Lee et al. (2018); thus, on average, a 1:1 ratio of CD4:CD8 cells in CAR-T therapy is more effective. Note that it is to be expected that some patients will benefit more from CD8 treatment than 1:1 treatment due to specific patient/tumor characteristics (i.e. high CD8 kill rate, more ineffective ratios of CD4 parameters:CD8 parameters, as in Table 2).

\subsubsection{Response and expansion behavior across CD4:CD8 ratios}

\begin{figure}[H]
    \centering
    \includegraphics[width=0.6\textwidth]{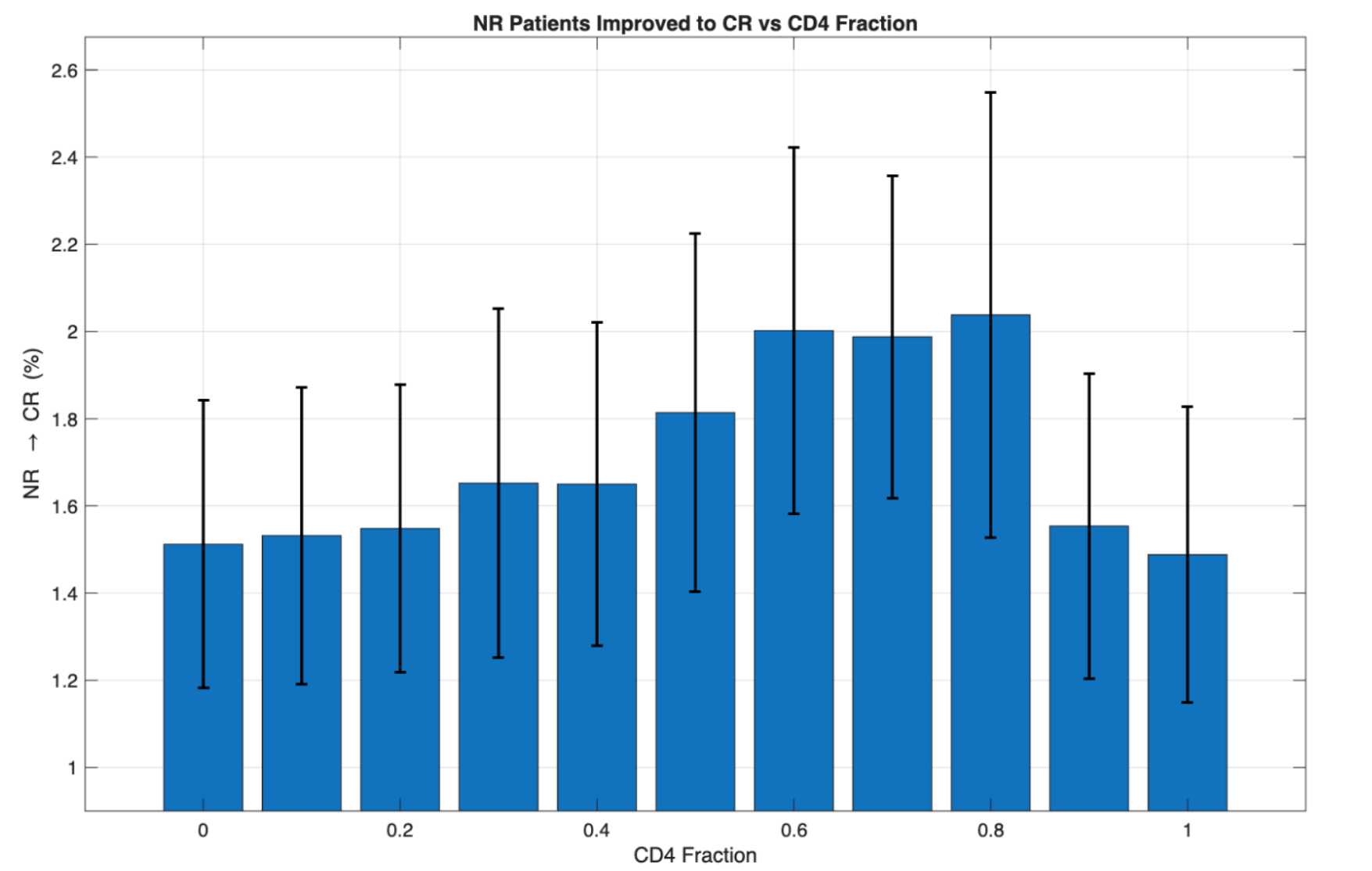} 
    \caption{\textbf{Fig. 5} Percentage of baseline non-responders rescued to complete response as a function of CD4 fraction in the CAR-T product. Results are shown for five independent trials of 5{,}000 virtual patients each, with newly generated virtual patient cohorts in every trial. Bars represent the mean rescue percentage, and error bars denote the standard deviation across trials}
    \label{fig:rescue_5trials}
\end{figure}

To further examine how CD4:CD8 composition influences treatment outcomes beyond the clinically motivated 1:1 case, we performed two complementary analyses that align more closely with experimentally reported metrics. First, we quantified the proportion of baseline non-responders (NR) that convert to complete response (CR) under CAR-T products with varying CD4 fractions, repeating this analysis across five independently generated virtual patient cohorts to assess sampling variability (Figure~\ref{fig:rescue_5trials}). On average, higher CD4 fractions were associated with increased NR$\rightarrow$CR rescue up to an intermediate range, reflecting the requirement for a balance between CD4$^+$ helper support and CD8$^+$ cytotoxic capacity; products dominated by either lineage alone were less effective. This qualitative trend is consistent with reported CAR-T expansion profiles, which show maximal expansion at intermediate CD4 fractions rather than at extreme compositions (Lee et al., 2022).

\begin{figure}[H]
    \centering
    \includegraphics[width=0.7\textwidth]{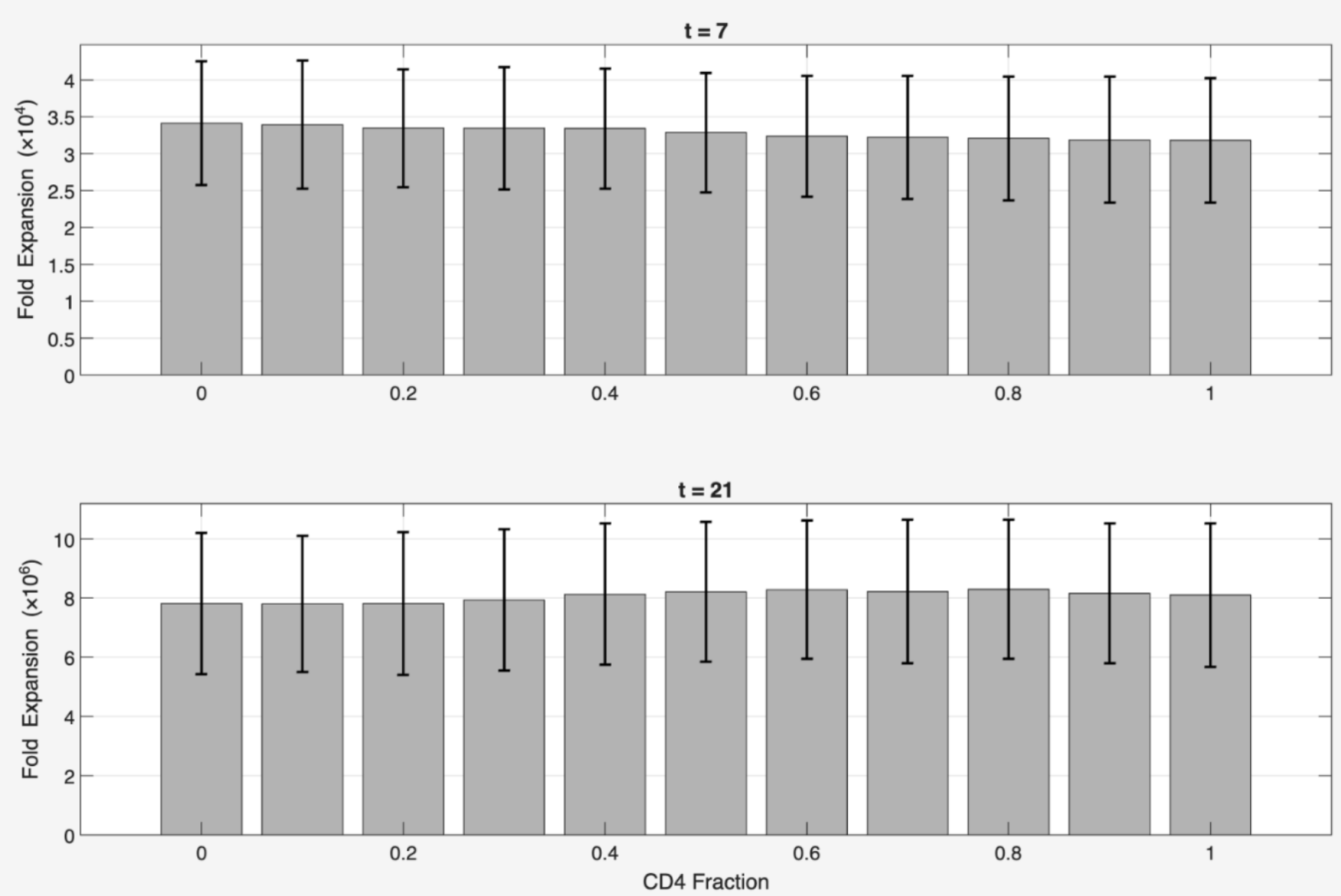} 
    \caption{\textbf{Fig. 6} CAR-T fold expansion as a function of CD4 fraction during the early and late phases of the initial expansion (“burst”) period. Shown is the mean fold expansion at $t = 7$ (top) and $t = 21$ (bottom) across simulated virtual patients for varying CD4 fractions in the infused CAR-T product. Fold expansion was quantified as the total CAR-T cell population evaluated at the indicated time point, normalized by the cumulative effective treatment exposure, computed as the area under the curve (AUC) of the DoseX state from infusion to the evaluation time. Error bars denote the standard deviation across patients. Results are averaged over five independent trials, each consisting of 5{,}000 independently generated virtual patients}
    \label{fig:expansion_5trials}
\end{figure}

Second, because the experimental study by Lee et al. primarily reported CAR-T expansion rather than response classifications, we directly examined CAR-T expansion dynamics as a function of CD4 fraction (Figure~\ref{fig:expansion_5trials}) over time. We computed the mean fold expansion of the total CAR-T population during the early ($t = 7$) and later ($t = 21$) phases of the post-infusion expansion burst. Early expansion was modestly higher for CD8-enriched products, consistent with the dominant cytotoxic role of CD8$^+$ effector cells, whereas at later times increased CD4 fractions supported slightly greater overall expansion. Notably, the observed trends depend on the time at which expansion is evaluated, and large error bars indicate pronounced inter-patient variability; accordingly, these results are interpreted as broadly qualitative rather than definitive.

\subsubsection{Response stratification under random CD4:CD8 CAR-T compositions}

The CD4:CD8 ratio in CAR-T therapy without a defined formulation can vary widely depending on the patient’s original lymphocyte composition. Galli et al. analyzed the CD4:CD8 ratios in such patients, and found that on average, complete responders had a CD4:CD8 ratio that was closer to 1:1 (smaller) than non-responders (Galli et al. 2023). To test our model against this finding, the virtual patients were treated with a random CD4:CD8 ratio (fraction of CD4 cells was randomly chosen from 0 to 1), and then categorized as CR or NR. It was found that the average fraction of cells that were CD4 in complete responders was 0.488 (standard deviation 0.268), and in non-responders it was 0.558 (standard deviation 0.301). This is in agreement with the data from Galli et al. (2023); on average, patients who respond to treatment  have a higher CD4:CD8 ratio (closer to 1:1).

\subsection{Parameter Selection for Practical Modeling} 

In theory, knowing the values of a patient's parameters would allow prediction of treatment outcomes for any CD4:CD8 ratio using the model. In practice, most of these parameters, although biologically meaningful, are impractical or impossible to measure before treatment. However, some of the most sensitive parameters (\( N \), \( k_{b2} \), \( k_{e}^{CD8} \), \( k_{b1} \), \( B50_{CD8} \)) may be approximated as follows. (Rather than proposing a complete clinical workflow, this subsection demonstrates, using well-established experimental assays and clinically accessible measurements, how a subset of the most influential parameters identified by sensitivity analysis can be approximated in practice and mapped onto the model. The goal is to provide a concrete proof-of-concept for measurement-informed modeling, not a finalized protocol for clinical deployment.) 

The parameter \( N \) (number of effector cell doublings during the proliferation burst) is difficult to directly measure before administration of treatment but can be estimated using current patient ranges and a measure of the proliferative capacity of T cells relative to other patients. For example, flow cytometry with CFSE (carboxyfluorescein succinimidyl ester), which binds to intracellular amines and whose brightness decreases as the cell divides, allows measurement of proliferation (Hawkins et al. 2007). A protocol would involve labeling CAR-T cells with CFSE, co-culturing them with CD19-expressing targets, measuring fluorescence dilution over 3-7 days, determining the percentage of cells in each generation, and comparing the average generation with other patients' values to estimate a value of \( N \). Alternatively, IL-2 levels in the supernatant post-stimulation with antigen can serve as a measure of activation, though this would require calibration against direct proliferation assays (CFSE dilution) to estimate \( N \). 

The parameter \( k_{b2} \) (antigen clearance rate) can be estimated using clearance rates of similarly sized proteins or with radiolabeled or tagged antigen if feasible. The parameter \( k_{e}^{CD8} \) (CD8 effector proliferation rate) can be estimated using CFSE dilution in CD8\textsuperscript{+} T cell subsets during in vitro stimulation. Additionally, \( k_{b1} \) (antigen generation rate) can be estimated indirectly using tumor burden, determined from blood work, and soluble antigen (CD19) levels in the blood. 

\( B50_{CD8} \) is measurable following production of CAR-T cells; in vitro antigen titration assays can be done using CAR-T cells stimulated with varying concentrations of soluble CD19 protein, then proliferation (via CFSE dilution) can be measured to determine the antigen concentration corresponding to half-maximal proliferation. Before CAR-T therapy, \( B50_{CD8} \) can be indirectly estimated using patient tumor burden and soluble CD19 levels from bloodwork, and by examining prior immune profiling of CD8\textsuperscript{+} T cell responses to CD19\textsuperscript{+} blasts, which together can indicate relative antigen sensitivity across patients.

\subsection{Neural Networks for Treatment Response Prediction} 

If the five aforementioned parameters can be measured, they can be used as inputs to the model, while all other parameters are set to typical patient values, allowing simulation of treatment outcomes. 

To assess the robustness of model-based response classification to parameter uncertainty, we introduced additive Gaussian noise to the subset of five key patient-specific parameters identified by sensitivity analysis (\(N\), \(k_{b1}\), \(k_{b2}\), \(k_{e}^{CD8}\), and \(B50_{CD8}\)). For each virtual patient, noisy parameter values were generated by sampling from normal distributions centered at the true parameter values, with standard deviations equal to 5\%, 10\%, 25\%, or 50\% of the biologically feasible parameter ranges reported by Kirouac et al.~(2023). These noise levels are not intended to model any specific experimental measurement process. Rather, they are designed to probe the limits of predictive performance under increasing uncertainty, including regimes in which parameter imprecision is large relative to biological variability (e.g.\ due to sparse, indirect, or noisy clinical measurements). This analysis therefore evaluates how sensitive mechanistic predictions are to parameter accuracy, and identifies conditions under which direct model-based classification may break down.
Each patient was simulated under multiple treatment scenarios (no treatment, and CAR-T cell therapies with varying CD4:CD8 ratios) using both the true and noisy parameter sets. We classified treatment outcomes (complete response vs. non-response) for both parameter sets and assessed the consistency between them to quantify the model’s robustness. 

\begin{table}[H]
\centering
\small
\renewcommand{\arraystretch}{1.3}
\setlength{\tabcolsep}{6pt}
\begin{tabular}{|c|c|c|c|c|}
\hline
\textbf{Noise (\%)} &
\textbf{Sensitivity (\%)} &
\textbf{Specificity (\%)} &
\textbf{Precision (\%)} &
\textbf{Accuracy (\%)} \\
\hline
5  & 84.1 & 62.4 & 69.1 & 73.3 \\
10 & 78.1 & 60.9 & 66.6 & 69.5 \\
25 & 77.3 & 58.2 & 64.9 & 67.8 \\
50 & 75.7 & 51.5 & 55.4 & 62.3 \\
\hline
\end{tabular}
\caption{\small Robustness of treatment response classification to parameter uncertainty. Average sensitivity (TPR), specificity (TNR), precision, and accuracy are reported across all tested conditions (No treatment; CD4 fraction = 0.0, 0.5, 0.8, and 1.0) as a function of increasing Gaussian noise applied to selected model parameters (5--50\% of parameter range). Results correspond to simulations of $n = 5000$ virtual patients.}
\label{tab:noise_levels_summary}
\end{table}

As parameter uncertainty increases, the mechanistic model’s ability to correctly classify treatment response (CR vs.\ NR) deteriorates rapidly (Table~\ref{tab:noise_levels_summary}). This degradation would be most consequential in regions of parameter space near any CR/NR decision boundary, where small errors can flip predicted outcomes. Notably, these are precisely the regimes in which predictive confidence is most clinically relevant: patients whose outcomes are sensitive to small changes in immune expansion or antigen dynamics are those for whom treatment selection and stratification are most challenging.
In the remainder of this section, we therefore focus on the noisiest setting (50\% of the biological parameter range) as a worst-case scenario to assess predictive limits under high uncertainty.

\begin{table}[H]
\centering
\small
\renewcommand{\arraystretch}{1.3}
\setlength{\tabcolsep}{6pt}
\begin{tabular}{|l|c|c|c|c|}
\hline
\textbf{Treatment} &
\textbf{Sensitivity (\%)} &
\textbf{Specificity (\%)} &
\textbf{Precision (\%)} &
\textbf{Accuracy (\%)} \\
\hline
NoTx          & 73.4 & 54.7 & 53.7 & 62.5 \\
CD4frac 0.0   & 77.0 & 49.6 & 56.5 & 62.2 \\
CD4frac 0.5   & 77.4 & 49.5 & 56.7 & 62.3 \\
CD4frac 0.8   & 76.1 & 50.6 & 55.5 & 62.1 \\
CD4frac 1.0   & 74.6 & 53.0 & 54.5 & 62.3 \\
\hline
\end{tabular}
\caption{\small Classification performance of the tumor-immune ODE model under parameter noise ($n=5000$) across CD4:CD8 treatment conditions, with Gaussian noise applied at 50\% of the biological parameter ranges reported by Kirouac et al. (2023). Sensitivity (true positive rate), specificity (true negative rate), precision (positive predictive value), and accuracy are reported.}
\label{tab:noise_sensitivity}
\end{table}

Metrics such as true positive rate (sensitivity) and true negative rate (specificity) were computed across all treatment conditions to summarize performance under this noisiest condition (Table~\ref{tab:noise_sensitivity}). The poor performance of the model under noise, particularly specificity and precision, indicate that parameter measurement noise would make it difficult to reliably predict treatment outcomes. To address this challenge, we implemented a feed-forward neural network to train the model on this noisy data. The goal was to leverage the network's ability to learn patterns from noisy inputs and improve predictive ability, thus mitigating the impact of measurement noise on the treatment outcome predictions.

We trained a multi-label feed-forward neural network to predict, for each virtual patient, whether the mechanistic model yields a complete response (CR) or non-response (NR) under five treatment conditions simultaneously (one output per treatment). Inputs were the patient’s noisy values for five model parameters; targets were the corresponding noiseless simulation outcomes (CR vs.\ NR). Each input variable was standardized by subtracting the training-set mean and dividing by the training-set standard deviation to place features on a comparable scale. Virtual patient parameters in this study are sampled independently across dimensions, so the neural network cannot exploit cohort-specific parameter correlations to implicitly denoise inputs. Consequently, improvements in predictive performance reflect recovery of genuine signal under Gaussian noise rather than overfitting to spurious correlation structure. (Identifying and incorporating realistic biological correlations between parameters remains an important direction for future empirical work.)

The network had three fully connected hidden layers (128, 64, and 32 units) with Rectified Linear Unit (ReLU) activations and 0.2 dropout to reduce overfitting. The final linear layer produced five independent logits (one per treatment); applying a sigmoid to each logit yields a CR probability that is thresholded to classify CR vs.\ NR. Optimization used AdamW (learning rate \(10^{-4}\), weight decay \(10^{-4}\)). Because several treatments are class-imbalanced, training used a focal loss (\(\gamma=2\)) with per-treatment class weights \(w_j=\#\text{neg}_j/\#\text{pos}_j\), which emphasizes harder or minority examples. 

Data were split into a held-out test set (20\%) and, from the remaining 80\%, a validation set (20\% of the remainder; 16\% overall) and a training set. Training ran for up to 200 epochs with early stopping based on validation balanced accuracy at a fixed 0.5 threshold, restoring the best checkpoint. After training, decision thresholds were calibrated per treatment by sweeping a grid \(t\in[0.01,0.99]\) on the validation set and selecting the threshold that maximized validation balanced accuracy. These validation-optimized thresholds were then applied once to the test set, and we report per-treatment sensitivity, specificity, precision, and accuracy (Table \ref{tab:nn_noise_sensitivity}).

\begin{table}[H]
\centering
\small
\renewcommand{\arraystretch}{1.3}
\setlength{\tabcolsep}{6pt}
\begin{tabular}{|l|c|c|c|c|}
\hline
\textbf{Treatment} &
\textbf{Sensitivity (\%)} &
\textbf{Specificity (\%)} &
\textbf{Precision (\%)} &
\textbf{Accuracy (\%)} \\
\hline
NoTx          & 66.0 & 66.5 & 66.3 & 66.3 \\
CD4frac 0.0   & 73.5 & 60.8 & 65.2 & 67.2 \\
CD4frac 0.5   & 72.5 & 60.7 & 64.9 & 66.6 \\
CD4frac 0.8   & 74.2 & 57.4 & 63.5 & 65.8 \\
CD4frac 1.0   & 77.3 & 54.8 & 63.1 & 66.0 \\
\hline
\end{tabular}
\caption{\small Classification performance of the neural network on the test set (20 \% of $n=5000$) across CD4:CD8 treatment conditions, with Gaussian noise applied at 50\% of the biological parameter ranges reported by Kirouac et al. (2023). Sensitivity (true positive rate), specificity (true negative rate), precision (positive predictive value), and accuracy are reported.}
\label{tab:nn_noise_sensitivity}
\end{table}

\begin{table}[H]
\centering
\small
\renewcommand{\arraystretch}{1.3}
\setlength{\tabcolsep}{8pt}
\begin{tabular}{|l|c|c|c|c|}
\hline
\textbf{Model} &
\textbf{Sensitivity (\%)} &
\textbf{Specificity (\%)} &
\textbf{Precision (\%)} &
\textbf{Accuracy (\%)} \\
\hline
Neural network & 72.7 & 60.0 & 64.6 & 66.4 \\
Naïve baseline & 64.8 & 35.2 & 64.9 & 50.1 \\
\hline
\end{tabular}
\caption{\small Average classification performance of the neural network compared with a naïve CR-rate baseline, averaged across the tested conditions (No treatment; CD4 fraction = 0.0, 0.5, 0.8, and 1.0) and trained on data with substantial Gaussian noise (standard deviation of 50\% of the biological parameter range). Average sensitivity (true positive rate), specificity (true negative rate), precision (positive predictive value), and accuracy are reported. Results correspond to performance on the test set (20\% of $n = 5000$ virtual patients).}
\label{tab:nn_vs_naive}
\end{table}

\begin{figure}[H]
\centering
\includegraphics[width=0.5\linewidth]{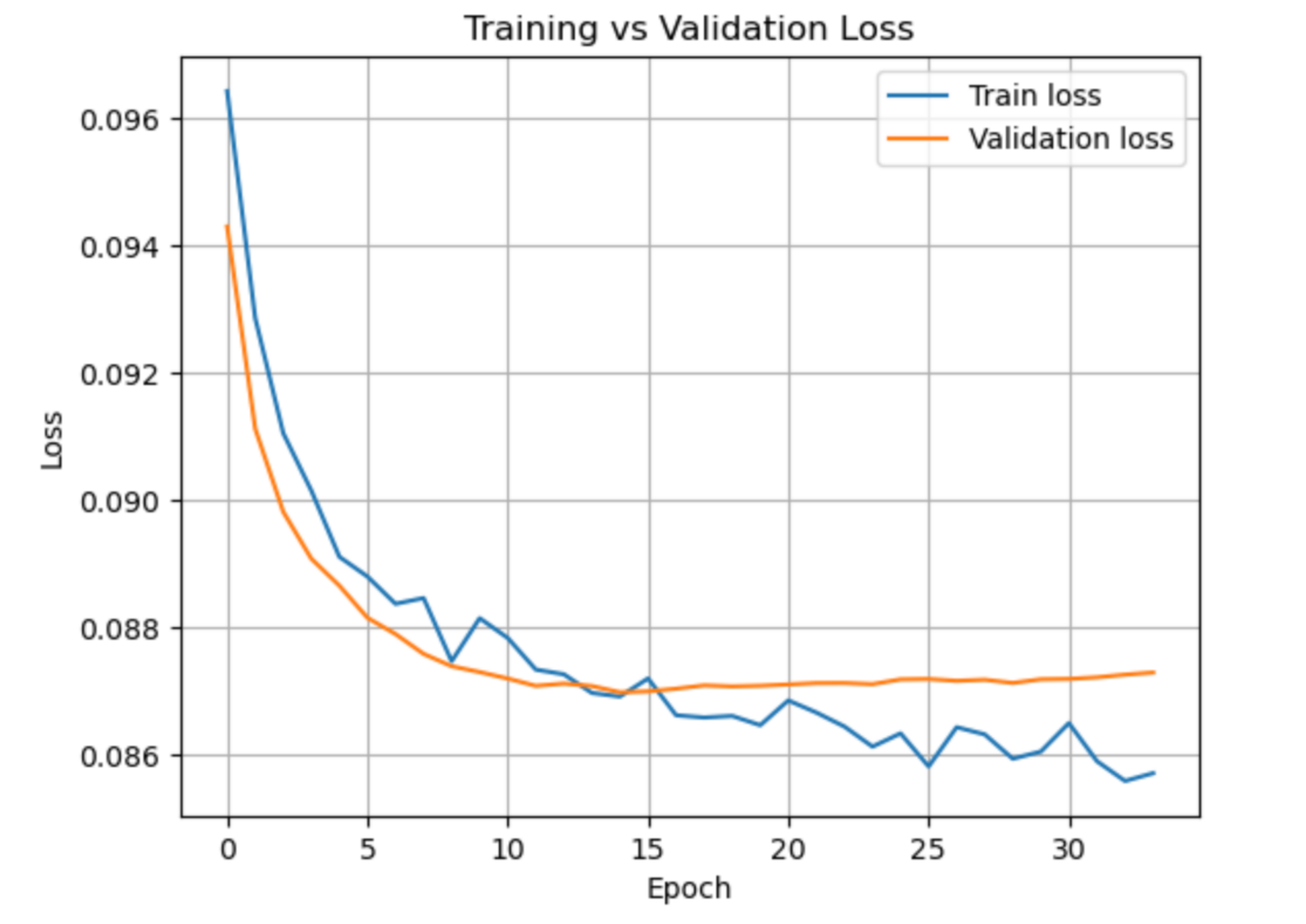}
\caption{\textbf{Fig. 7} Training and validation loss per epoch. Lower validation loss is better (indicates generalization), and the best model is chosen at the minimum validation loss via early stopping}
\label{fig:loss_curves}
\end{figure}

\begin{figure}[H]
\centering
\includegraphics[width=0.5\linewidth]{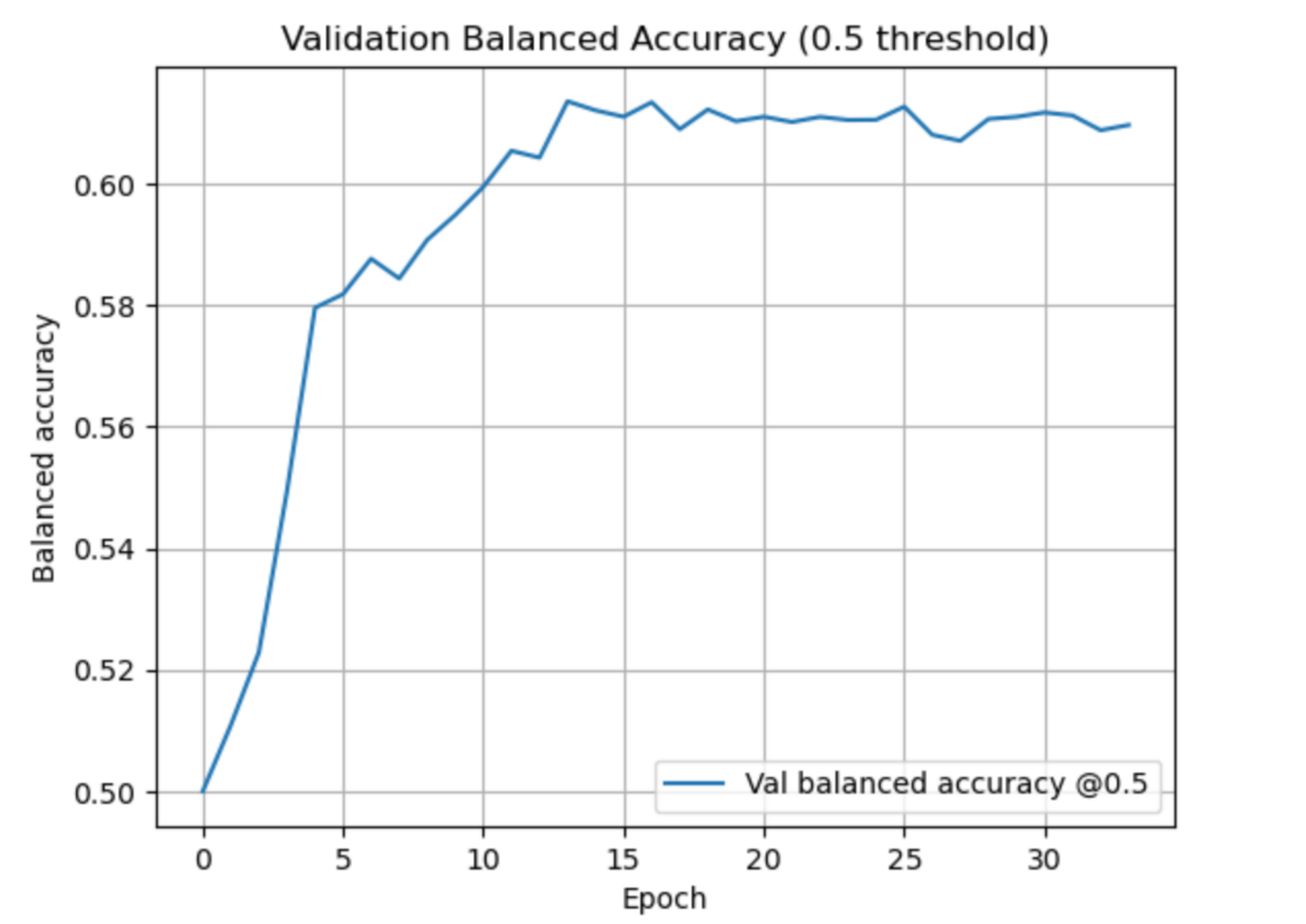}
\caption{\textbf{Fig. 8} Validation balanced accuracy per epoch, computed as the average of sensitivity and specificity. The rise followed by a plateau suggests the model learns useful signal and then converges}
\label{fig:balacc_curve}
\end{figure}

As epochs increase, the training loss steadily decreases, while the validation loss drops at first and then plateaus; stopping training near the point where the validation loss plateaus (Fig.~\ref{fig:loss_curves}) helps prevent overfitting; i.e., it keeps the model from learning dataset-specific noise and preserves generalization to unseen data. In Fig.~\ref{fig:balacc_curve}, the validation balanced accuracy rises quickly over the early epochs and then levels off, indicating that the network is learning useful signal and then converging. The plateau (with only small fluctuations) suggests diminishing returns from additional training rather than continued improvement, consistent with a stable model that generalizes to held-out data. Because balanced accuracy averages sensitivity and specificity, the upward trend reflects gains in both detecting responders and correctly rejecting non-responders despite class imbalance. Consistent with this, comparison to a naïve CR-rate baseline (Table~\ref{tab:nn_vs_naive}) shows that the network’s performance gains are not driven by learning a global response bias, but by improved discrimination, most notably higher specificity and overall accuracy. Because the naïve classifier predicts CR at a fixed rate matched to the empirical prevalence, its precision is necessarily close to its sensitivity and reflects population frequency rather than true discriminative power. The superior performance of the neural network, therefore, indicates recovery of predictive signal beyond population-level statistics.

Although the neural network exhibits slightly reduced sensitivity relative to direct model-based classification in some treatment conditions, it substantially improves specificity, precision, and overall accuracy under high parameter uncertainty. This reflects a more conservative but more reliable classification strategy, reducing false-positive predictions while maintaining comparable true-positive detection. Consequently, the neural network achieves markedly higher balanced accuracy, indicating superior performance in the presence of severe parameter noise. Importantly, this shift toward higher precision and specificity may be clinically valuable, since false-positive responder predictions can lead to overly optimistic treatment selection and misguided stratification in the settings where uncertainty is highest.

\begin{figure}[H]
    \centering
    \begin{subfigure}[b]{0.8\linewidth}
        \centering
        \includegraphics[width=\linewidth]{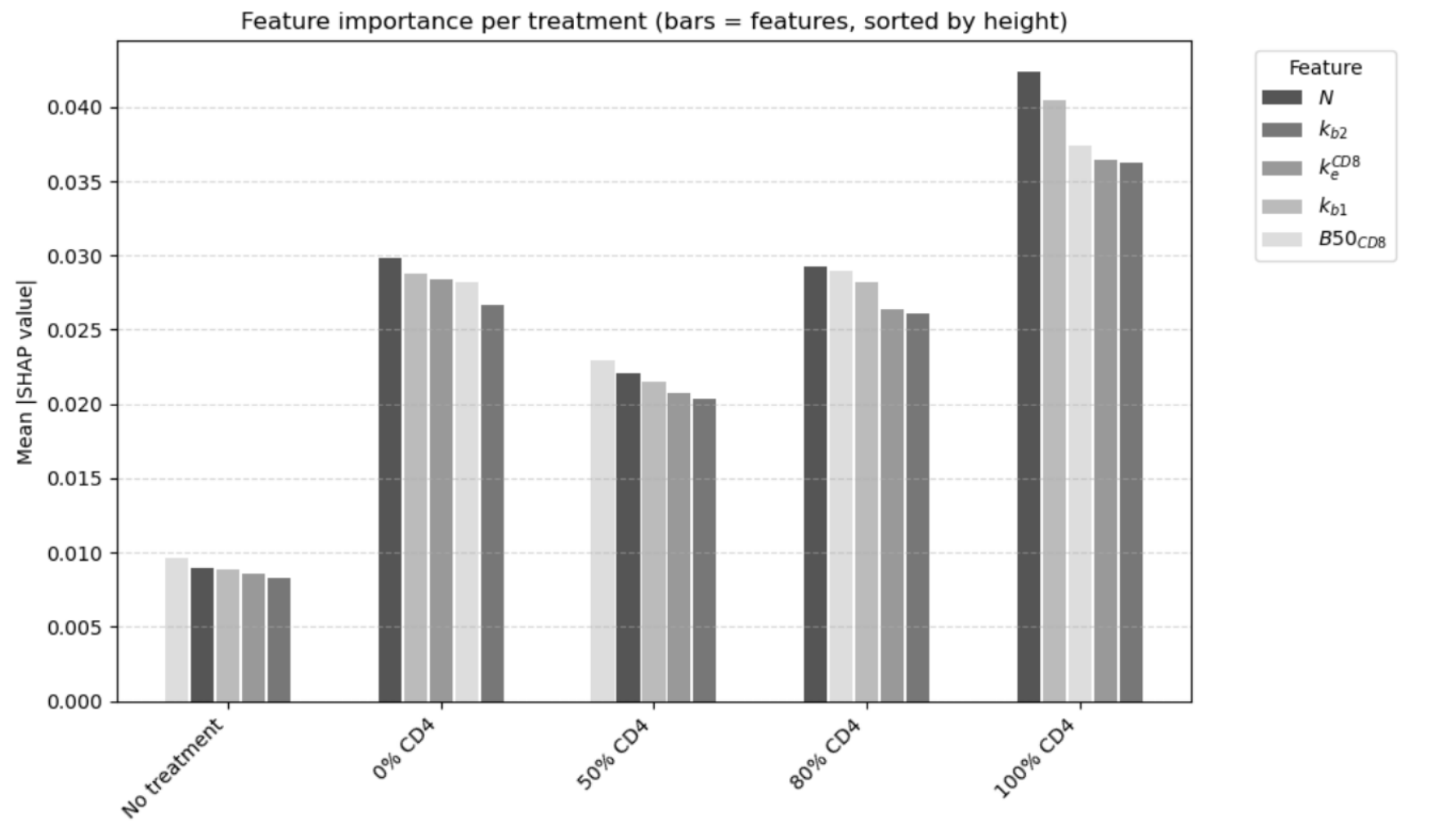}
        \caption{Individual SHAP results for each treatment scenario. Bar colors (dark to light) follow the Sobol sensitivity ranking, with darker bars indicating more influential parameters, highlighting their relative contributions to the model output}
        \label{fig:shap_individual_a}
    \end{subfigure}
    \hfill
    
    \begin{subfigure}[b]{0.6\linewidth}
        \centering
        \includegraphics[width=\linewidth]{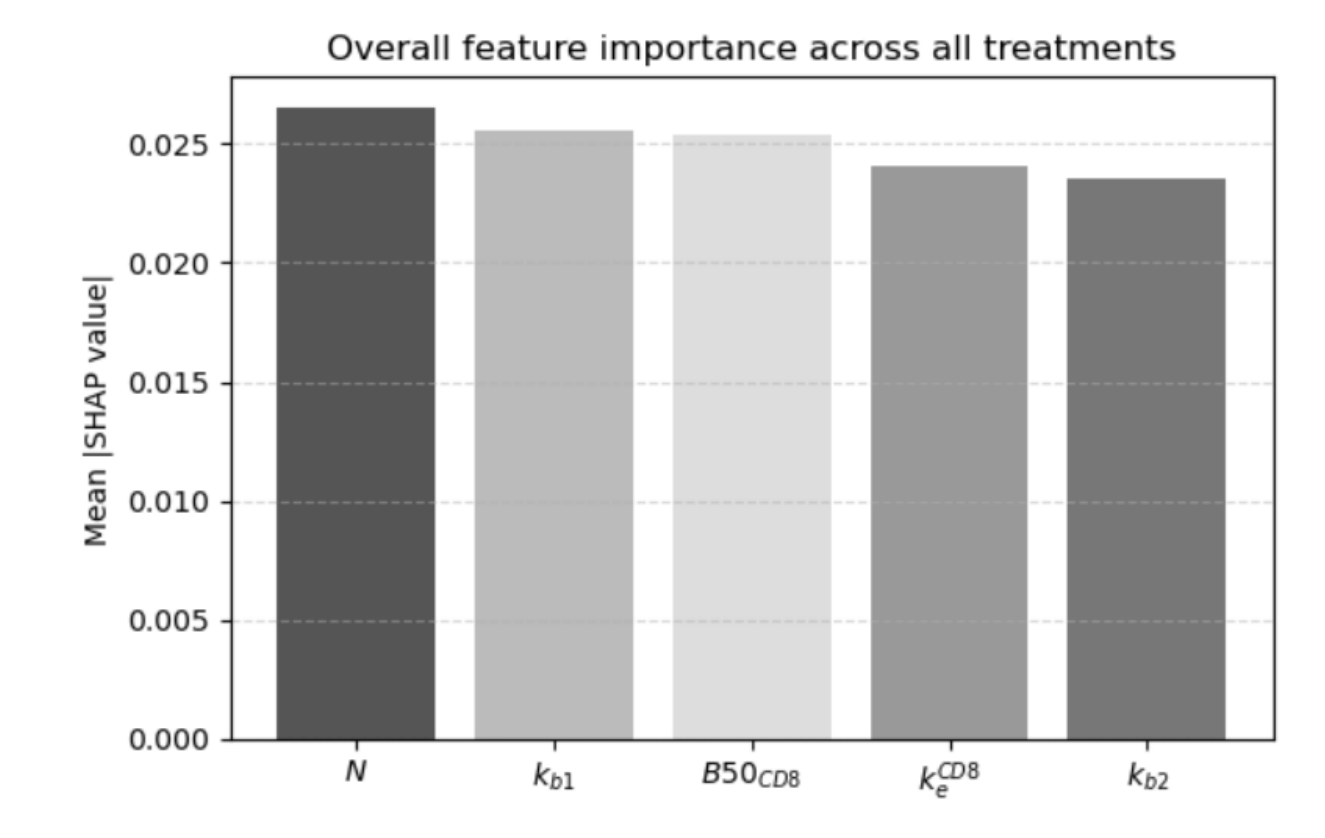}
        \caption{Combined SHAP results. Averages were taken across all treatments}
        \label{fig:shap_combined_b}
    \end{subfigure}

    \begin{subfigure}[b]{0.6\linewidth}
        \centering
        \includegraphics[width=\linewidth]{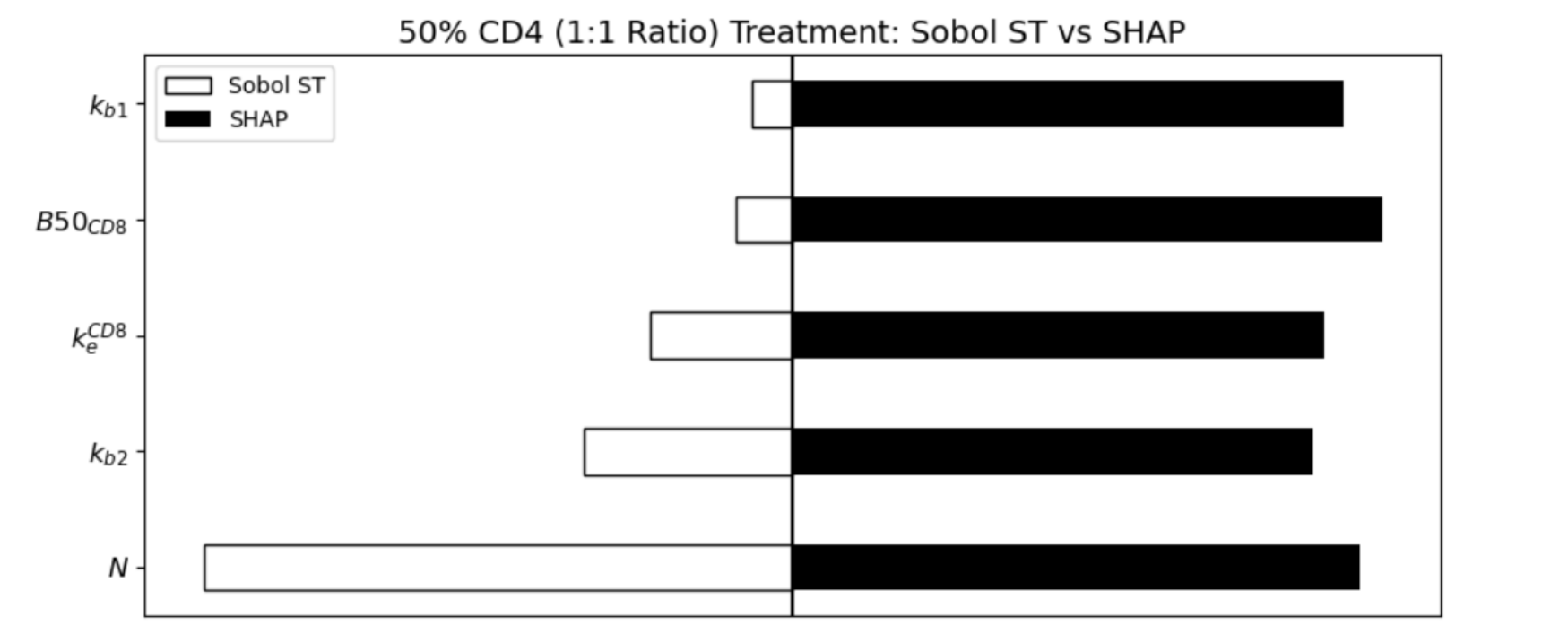}
        \caption{Parameter importance for 50\% CD4 treatment: Sobol ST indices (left) and SHAP values (right), normalized to each side’s largest bar. Sobol ST indices were similar across treatments, so 1:1 CD4:CD8 treatment is shown as a representative}
        \label{fig:shap_compare_c}
    \end{subfigure}

    \caption{\textbf{Fig. 9} SHAP analysis of the model}
    \label{fig:shap_overall}
\end{figure}

We also used SHAP analysis to interpret feature importance across treatments (Figure~\ref{fig:shap_individual_a}). Consistent with prior sensitivity analyses, variability in $N$ was the most influential factor overall (Figure~\ref{fig:shap_combined_b}). The relative importance of the remaining parameters ($k_{b2}$, $k_{e}^{CD8}$, $k_{b1}$, $B50_{CD8}$) varied slightly across local, PRCC, and Sobol analyses, which is reflected in the SHAP ranking not exactly matching the Sobol order (Figure~\ref{fig:shap_compare_c}). This indicates that while $N$ largely drives system behavior, interactions among the other parameters likely contribute nonlinearly to treatment outcomes. These results illustrate the value of combining mechanistic modeling with machine learning to better understand complex parameter effects under uncertainty.

\section{Discussion}

In this study, we extended the CAR-T cell therapy model proposed by Kirouac et al., which models antigen-regulated transitions between memory, effector, and exhausted T cell states, to explicitly model CD4\textsuperscript{+} and CD8\textsuperscript{+} T cell subtypes and their distinct roles in tumor-immune dynamics. As far as we know, our model is the first to differentiate between CD4\textsuperscript{+} and CD8\textsuperscript{+} CAR-T cells, allowing investigation of the synergistic interactions between these subtypes and their impact on treatment outcomes. Importantly, we do not view our extended model as a replacement for the original Kirouac et al.\ framework, but rather as a mechanistically interpretable extension that preserves its structure while expanding its biological scope.

We use the extended model to investigate the effects of varying CD4:CD8 ratios on CAR-T cell efficacy and tumor control. In particular, the 1:1 CD4:CD8 ratio was examined because defined-composition CAR-T products with approximately equal CD4\textsuperscript{+} and CD8\textsuperscript{+} fractions have been investigated clinically and shown to yield improved efficacy relative to CD8-only formulations (Turtle et al., 2016). Our results demonstrate that a balanced 1:1 ratio of CD4\textsuperscript{+} to CD8\textsuperscript{+} CAR-T cells leads to more effective tumor elimination and CAR-T cell expansion on average (compared with a nonspecific CD4:CD8 ratio) but not always. As such, an important next step would be identification of the patient factors that determine whether a 1:1 CD4:CD8 treatment ratio or an unspecified ratio yields better therapeutic outcomes. Machine learning can be employed to analyze virtual patient simulation results to identify these factors. Identification of patient subgroups that are most likely to benefit from a specific CD4:CD8 ratio can enhance patient stratification and allow more personalized treatment. However, predictability of treatment response is dependent on limitations in quantifying patient-specific differences. 

This highlights the distinction between mechanistic insight and predictive utility. Although the extended Kirouac et al. framework exhibits nonlinear outcome transitions under specific parameter variations, these behaviors depend strongly on patient-specific parameter combinations and are not universal; therefore, we used sensitivity analyses to systematically identify the parameters most influential for clinically relevant outcomes. We identified five such key parameters, and explored the effects of noise in patient-specific measurements of these parameters. Importantly, the noise levels considered here are intended as stress tests of predictive reliability rather than realistic models of experimental measurement error. While these noise levels may exceed typical experimental variability, substantial uncertainty can arise when parameters are inferred indirectly or estimated from sparse clinical data.

We demonstrate that although noise in parameter measurements hinders direct predictions of treatment outcomes using the mathematical model, machine learning provides a potential solution for identifying possible treatment responders. We trained a neural network on noisy parameter sets and corresponding simulation outcomes to predict virtual patient responses, achieving improved overall performance for direct predictions from the mathematical model. It is important to note, however, that this recovery is partial and does not eliminate the need for accurate parameter measurement; rather, it highlights regimes where data-driven methods can complement mechanistic models when uncertainty is unavoidable. Additionally, the agreement between SHAP and mechanistic sensitivity analyses strengthens confidence in the model’s predictions and illustrates the complementarity of mechanistic and machine learning approaches. 

One limitation of this model, however, is that it does not include cytokine dynamics. Cytokines are a diverse group of signaling proteins that play a crucial role in the immune system by regulating the abundance and distribution of various immune cell populations, and are thus integral to the mechanism of CAR-T therapy. Importantly, CD4\textsuperscript{+} T cells rely heavily on cytokine signaling for much of their helper functions, including activation of CD8\textsuperscript{+} T cells. In particular, IFN-$\gamma$ and IL-12 are cytokines that are necessary for maintaining CAR-T cytotoxicity (Boulch et al. 2021), making their inclusion in the model a valuable next step. Our model does implicitly account for some of these cytokine-mediated interactions, particularly through the modulation of CD8\textsuperscript{+} T cell killing by CD4\textsuperscript{+} T cells, it does not explicitly capture the dynamics of cytokines involved in tumor-immune interactions. As such, a key future direction for this model is the inclusion of cytokine dynamics. 

With the inclusion of cytokines in the model, it is also possible to study side effects, particularly those related to cytokines such as IL-6 and IFN-$\gamma$, which are central to the pathogenesis of cytokine release syndrome (CRS) and neurotoxicity observed in CAR-T cell therapy (Boulch et al. 2023; Gust et al. 2020). CRS is a common side effect of CAR-T therapy caused by overactivation of the immune system; it presents with fever, hypotension, and elevated levels of cytokines, including IL-6 and IFN-$\gamma$. This can lead to more severe side effects, including neurotoxicity. Studies have shown that patients who develop neurotoxicity after CAR-T therapy have significantly higher levels of IL-6 and IFN-$\gamma$ early post-infusion, and these cytokine levels may serve as biomarkers for identifying those at risk of severe side effects (Turtle et al. 2016). Including these cytokine dynamics in the model would allow for a more comprehensive understanding of how CAR-T therapy not only targets tumor cells but also triggers immune-related toxicities. Specifically, by modeling the release and impact of IL-6 and IFN-$\gamma$, the model could help predict the likelihood of CRS and neurotoxicity, allowing for better patient stratification and earlier interventions to mitigate these side effects. Moreover, it would be beneficial to explore how these cytokines influence tumor-immune dynamics in conjunction with CAR-T efficacy, potentially balancing the therapeutic benefits against the risks of immune-related toxicity.

Additionally, our model assumes a simplified representation of tumor-immune dynamics, neglecting factors such as tumor heterogeneity, immune suppression, and interactions with other immune cells. Incorporating these factors would enable more detailed and complex modeling, potentially allowing for the identification of other key drivers in CAR-T therapy and improving predictive power. However, this would also complicate the analysis, as it would introduce more variables and necessitate more accurate and comprehensive data, which may be challenging to obtain. Ultimately, finding the right balance between complexity and simplicity is key to creating the most useful model. By starting with a solid foundation and adding essential elements, such as the inclusion of CD4\textsuperscript{+} and CD8\textsuperscript{+} interactions, we have taken a meaningful step towards more accurate predictions in CAR-T therapy.

Another key limitation of the current model is the lack of sufficient patient data, which is essential for refining the model structure and parameterizing it accurately. With more clinical data, we would be able to make more robust predictions for individual patients. As explored, machine learning can help overcome the challenge of noisy or incomplete clinical data, allowing for more accurate predictions even when data is uncertain, thereby improving clinical decision-making. As such, using machine learning algorithms trained on a larger set of patient data could allow predictions of patient-specific responses to CAR-T therapy. This approach, akin to systems like LORIS (Chang et al. 2024), would improve patient stratification, ensuring that those most likely to benefit from the treatment are identified. 

\section{Statements and Declarations}
This work was supported by the Natural Sciences and Engineering Research Council of Canada (NSERC). MK gratefully acknowledges this support. The authors have no competing interests to declare that are relevant to the content of this article. 

\section{Data Availability}
This study did not generate new experimental data. All data used for simulations were obtained from previously published sources, as cited in the manuscript. Simulation code and virtual patient data are available from the corresponding author upon reasonable request.

 \nocite{*}
\printbibliography
\end{document}